%% file: ASPIRE_manu_clean_draft.tex
\documentclass{article}
\usepackage{amsmath}
\usepackage{amssymb}
\usepackage[margin=1in]{geometry}

\usepackage{hyperref}
\usepackage{natbib}
\usepackage{xcolor}
\hypersetup{
    colorlinks,
    linkcolor={red!50!black},
    citecolor={blue!50!black},
    urlcolor={blue!80!black}
}

\usepackage{pgf,tikz}
\usetikzlibrary{arrows.meta,shapes.arrows,shapes.geometric,shapes.multipart,decorations.pathmorphing,positioning,shapes.swigs}

\usepackage{authblk}

\usepackage{multirow}
\usepackage{graphicx}
\usepackage{caption}
\usepackage{subcaption}
\usepackage{booktabs}
\usepackage{placeins}
\usepackage{colortbl}
\usepackage{tabularx}
\usepackage{makecell}
\usepackage{float}
\usepackage{tcolorbox}

\newcommand{\Bern}{\text{Bern}}
\newcommand{\expit}{\text{expit}}
\newcommand{\MVN}{\text{MVN}}
\newtheorem{identity}{Proposition}

\title{Estimating the average treatment effect in cluster-randomized trials with misclassified outcomes and non-random validation subsets}

\author[1]{Dane Isenberg\thanks{dane.isenberg@pennmedicine.upenn.edu}}
\author[1]{Nandita Mitra}
\author[2]{Steven C. Marcus}
\author[3]{Rinad S. Beidas}
\author[1]{Kristin A. Linn}
\affil[1]{Department of Biostatistics, Epidemiology and Informatics, University of Pennsylvania, Philadelphia, PA, USA}
\affil[2]{University of Pennsylvania School of Social Policy and Practice, Philadelphia, PA, USA}
\affil[3]{Department of Medical Social Sciences, Northwestern University Feinberg School of Medicine, Chicago, Illinois, USA}
\date{}

\begin{document}

\maketitle

\begin{abstract}
Randomized trials are viewed as the benchmark for assessing causal effects of treatments on outcomes of interest. Nonetheless, challenges such as measurement error can arise that undermine the standard causal assumptions for randomized trials. In ASPIRE, a cluster-randomized trial, pediatric primary care clinics were assigned one of two treatment arms aimed at promoting clinician delivery of a secure firearm program to parents during well-child visits. Accordingly, a key outcome of interest is parent receipt of
the program at each visit. Program delivery was documented by clinicians in patients’ electronic health
records for all visits, but these clinician reports are only a proxy measure for the parent receipt
outcome. Parents were also surveyed to report directly on their receipt of the program after their child’s
appointment; however, only a small subset of them completed the survey. Motivated by the ASPIRE trial, we develop a causal inference framework for randomized data where a binary outcome is subject to misclassification through silver-standard measures (clinician reports), and gold-standard measures (parent reports) are available for an internal validation subset but were not randomly sampled. Here, we propose a method for identifying the average treatment effect (ATE) that addresses the risk of bias due to imperfect outcome measurement and non-random validation selection, even in a setting such as ASPIRE where the outcome (parent-perceived delivery) directly impacts selection propensity (parent willingness to respond to a survey). We show that estimation of the ATE relies on specifying the relationship between the gold- and silver-standard outcome measures in the validation subset, which we allow to depend on treatment and covariates, marking a critical extension of the previous literature. Additionally, the clustered design is reflected in our causal assumptions and in our cluster-robust approach to estimation of the ATE. Simulation studies demonstrate acceptable finite-sample operating characteristics of our ATE estimator, supporting its application to the ASPIRE trial.
\end{abstract}

\section{Introduction}
Randomized trials are designed to satisfy core assumptions that support causal interpretations of estimated treatment effects. For a randomized trial comparing two treatment arms, a common treatment effect estimand of interest is the average treatment effect (ATE), the expected difference in potential outcomes across the two treatments. Under an ideal trial, the ATE is directly identified with the corresponding difference in observed means. However, if outcomes are measured with error, this identification may not hold, and a simple difference in means estimator using these measures may not target the ATE without bias \citep{carroll2006measurement}. Researchers often seek to collect accurate outcome measures from an internal subset of individuals in the study to help correct for measurement error, but it may not be feasible to obtain a representative subset, introducing the potential for selection bias in estimating the ATE for the full cohort \citep{hernan2004structural}. In this work, we codify a causal inference framework and propose an estimator for the ATE that addresses a combination of challenges associated with misclassified binary outcomes and a non-random internal validation subset.  This set of challenges arises in our motivating application,
a cluster-randomized trial referred to as ASPIRE\footnote{Adolescent and Child Suicide Prevention in Routine Clinical Encounters} \citep{beidas2024implementation,beidas2021study}. 

For the ASPIRE study, researchers examined two strategies to ensure that clinicians delivered an evidence-based firearm safety program, SAFE Firearm \citep{davis2021adapting,hoskins2021mixed,barkin2008office}, for parents or caregivers (hereafter, parents) during standard well-child visits. Full delivery of the program involved two components: counseling parents on safe firearm storage and offering them free cable firearm locks. The first strategy, denoted ``nudge", integrated a new electronic health record (EHR) documentation template into the standard well-child visit workflow, which reminds clinicians to administer SAFE Firearm and  to record whether they delivered the full program. The second strategy, called ``nudge+", combined the nudge protocol with clinic-level facilitation to provide clinicians with support in implementing SAFE Firearm. Because SAFE Firearm is directed at parents and their firearm storage practices, we can compare the effectiveness of the two strategies (on delivery) by examining their impact on parent receipt of the program at standard well-child visits. In ASPIRE, clinicians (self-)reported on their delivery of SAFE Firearm in the EHR documentation at all visits. The availability of these clinician reports is appealing, but they are only a proxy for measuring parent receipt of the program. Therefore, we refer to clinician reporting as a silver-standard measure \citep{yang2023machine,li2024multisource} of the parent receipt outcome because it may be subject to misclassification. In this study, researchers also surveyed parents after their child's appointment to inquire directly about whether their child's clinician had delivered both components of the program. Their responses serve as the gold-standard measure of the parent receipt outcome. However, there was a relatively low response rate, and the respondents were not randomly selected.
Here, we describe
how to leverage both the silver-standard clinician reporting and the gold-standard parent reporting to target
the ATE comparing nudge+ to nudge on parent receipt of SAFE Firearm.

Although ASPIRE is a randomized trial, naive approaches for estimating the ATE for the parent receipt outcome using clinician reports can lead to misclassification bias \citep{carroll2006measurement,althubaiti2016information}. Additionally, estimation using only parent reports may be vulnerable to selection bias \citep{hernan2004structural}. To address these potential issues, we first provide a general causal representation, using potential outcomes, of how the measurement error \citep{hernan2009invited} and selection mechanisms \citep{hernan2004structural,lu2022toward,mathur2025simple} operate in studies like ASPIRE, employing the aid of single world intervention graphs \citep{richardson2013single}. From this representation, we propose a set of identification assumptions that enable valid identification of the ATE with observed data from a cluster-randomized trial. We build on  \cite{shu2019causal}, which addressed outcome misclassification when targeting the ATE (drawing their motivation from \cite{braun2016using}), and \cite{shen2025integrating}, which adapted the former approach to settings with non-random selection of the validation subset.

Characterizing the relationship between silver-standard and gold-standard outcomes vis-\`{a}-vis the selection mechanism is fundamental to identification and estimation of the ATE. We consider a few unexplored complexities regarding this relationship, which arise in the ASPIRE trial. First, we allow the rates of (mis)classification to depend on treatment since treatment (i.e., exposure to nudge or nudge+) may impact both parent-perceived delivery of the program as well as clinician documentation of it. We also consider that baseline covariates, such as patient or site-specific characteristics, may influence how silver-standard outcomes relate to gold-standard outcomes for which we specify a \textit{classification model}. Moreover, since outcome occurrence temporally precedes selection (i.e., parents' receipt or nonreceipt of SAFE firearm precedes their documentation via survey response), we consider that the outcome itself affects the selection mechanism--a causal pathway that may invalidate standard inverse probability of selection weighting methods among the gold-standard data \citep{daniel2012using,seaman2013review,li2013weighting}. In this case, we leverage that silver-standard outcomes do not impact selection directly (since parents are agnostic to clinician reporting) to estimate the ATE. Lastly, we address the fact that treatments were randomly assigned to pediatric primary care clinics as opposed to individuals. The cluster-randomized design is built into our causal assumptions and accounted for in our variance estimation to avoid anti-conservative inference \citep{murray2004design,turner2017review2}. To accomplish the latter, we construct our estimator as a solution to cluster-indexed unbiased estimating equations, which permits desirable asymptotics and a cluster-robust sandwich variance estimator per M-estimation theory \citep{stefanski2002calculus,iverson1989effects,liang1986longitudinal,huber1967,huber1964robust}.

We have organized the remaining sections as follows. In Section~\ref{sec:estimand}, we define notation, describe causal mechanisms and assumptions, and identify the ATE. In Section~\ref{sec:estimation}, we present an estimator for the ATE with a general classification model that can include covariates but reduces to a non-parametric estimator when covariate effects are null. We also present its cluster-robust variance estimator using an M-estimation framework. In Section~\ref{sec:simstudy}, we conduct a simulation study to assess the finite-sample operating characteristics of our estimation approach, where we include a causal relationship between potential gold-standard outcomes and selection in data generation. We also compare our estimator to three standard alternatives that make critical simplifying assumptions. In Section~\ref{sec:application}, we apply our methods to the ASPIRE trial where we estimate the ATE for the effect of nudge+ versus nudge on the parent receipt outcome. Finally, we provide a discussion and future considerations in Section~\ref{sec:discussion}. To facilitate implementation of the proposed methods, we provide a companion \texttt{R} package, \texttt{mesbestim}, available at \url{https://github.com/abcdane1/mesbestim}.

\section{Targeting the Average Treatment Effect}
\label{sec:estimand}

\subsection{Notation and Setup}

Let $i=1,...,m$, represent the cluster index, and let $j=1,...,n_i$ represent the individual index within each cluster $i$. Therefore, $m$ is the total number of clusters, $n_i$ is the number of individuals in cluster $i$, and $N=\sum_{i=1}^m n_i$ is the total number of individuals. Since we have cluster-level randomization, we define the treatment indicator as $A_i \in \{0,1\}$; as per the setting of ASPIRE, we allow treatment to represent a non-pharmacological intervention. 
We use $Y_{ij}^{*} \in \{0,1\}$ to represent the silver-standard outcome values, which are prone to classification error, and $Y_{ij} \in \{0,1\}$ to represent the gold-standard outcome values, which are only measured for a selected subset of individuals. We will encode an individual's selection by the variable $V_{ij} \in \{0,1\}$, where $V_{ij}=1$ signifies that their outcome $Y_{ij}$ is recorded, and the number of selected individuals in each cluster is given by $n^v_i=\sum^{n_i}_{j=1}V_{ij}$. 

We define the target estimand, the ATE, in terms of individuals' potential outcomes, that is, individuals' outcomes had they in fact been assigned via their cluster to each of the treatments. We denote an individual's pair of potential outcomes as $Y_{ij}(a) \in \{0,1\}$ for $a \in \{0,1\}$, and we define the ATE, $\tau$, as
\begin{align}
\begin{split}
\tau&:=E\left\{Y_{ij}(1)-Y_{ij}(0)\right\}\\
&=P(Y_{ij}(1)=1)-P(Y_{ij}(0)=1)=\mu(1)-\mu(0).
\end{split}
\end{align}
Because each individual is assigned only to the treatment randomized to their cluster, at most one of their potential outcomes is observable if they are also in the validation subset. Next, we provide identification assumptions to allow the ATE to be represented in terms of observable data.

\subsection{Identification}

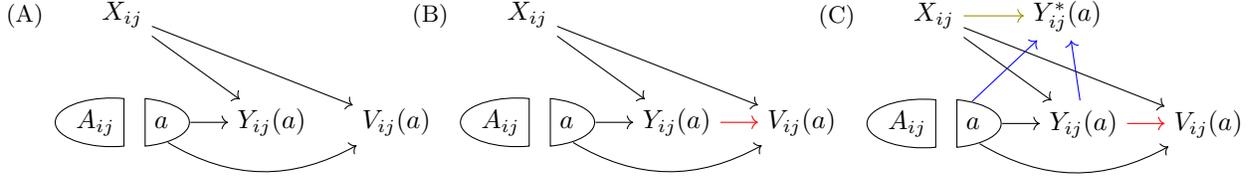
\begin{figure}[t]

\captionsetup{labelfont=bf,justification=raggedright,singlelinecheck=false}
    
    \centering
    \begin{minipage}{0.32\textwidth}
        \raisebox{9\height}{\small (A)}
        \begin{tikzpicture}
        \node[name=a,shape=swig vsplit]{
        \nodepart{left}{$A_{ij}$}
        \nodepart{right}{$a$}};
        \node[name=x, above=8mm of a]{$X_{ij}$};
        \node[name=y, right=5mm of a]{$Y_{ij}(a)$};
        \node[name=v, right=5mm of y]{$V_{ij}(a)$};
        \draw[->](a) to (y);
        \draw[->](x) to (y);
        \draw[->](x) to (v);
        \draw[->,bend right](a) to (v);
        \end{tikzpicture}
    \end{minipage}
    \begin{minipage}{0.32\textwidth}
    \raisebox{9\height}{\small (B)}
    \begin{tikzpicture}
    \node[name=a,shape=swig vsplit]{
    \nodepart{left}{$A_{ij}$}
    \nodepart{right}{$a$}};
    \node[name=x, above=8mm of a]{$X_{ij}$};
    \node[name=y, right=5mm of a]{$Y_{ij}(a)$};
    \node[name=v, right=5mm of y]{$V_{ij}(a)$};
    \draw[->](a) to (y);
    \draw[->,red](y) to (v);
    \draw[->](x) to (y);
    \draw[->](x) to (v);
    \draw[->,bend right](a) to (v);
    \end{tikzpicture}
    \end{minipage}
    \begin{minipage}{0.32\textwidth}
    \raisebox{9\height}{\small (C)}
\begin{tikzpicture}
    \node[name=a,shape=swig vsplit]{
    \nodepart{left}{$A_{ij}$}
    \nodepart{right}{$a$}};
    \node[name=x, above=8mm of a]{$X_{ij}$};
    \node[name=y, right=5mm of a]{$Y_{ij}(a)$};
    \node[name=v, right=5mm of y]{$V_{ij}(a)$};
    \node[name=ys, right=8mm of x]{$Y^*_{ij}(a)$};
    \draw[->](a) to (y);
    \draw[->,red](y) to (v);
    \draw[->](x) to (y);
    \draw[->](x) to (v);
    \draw[->,bend right](a) to (v);
    \draw[->,olive](x) to (ys);
    \draw[->,blue](y) to (ys);
    \draw[->,blue](a) to (ys);
\end{tikzpicture}
    \end{minipage}
    \caption{A set of three single world intervention graphs (SWIGs) for reference. $X_{ij}$ are covariates for an arbitrary individual $i$ in cluster $j$, which can be cluster-based or individual characteristics. $Y_{ij}(a)$, $Y^*_{ij}(a)$, and $V_{ij}(a)$ are potential gold-standard, silver-standard, and selection outcomes under treatment $A_{ij}=A_i=a \in \{0,1\}$ for an arbitrary individual $i$ in cluster $j$. SWIG (A): condition on $X_{ij}$ without internal or net-external bias, SWIG (B): condition on $X_{ij}$ without internal bias but with the potential for net-external bias, and SWIG (C) : condition on $X_{ij}$ and leverage the $Y^*_{ij}(a)$-$Y_{ij}(a)$ relationship to eliminate internal or net-external bias (as in ASPIRE). We will assume that it is not possible to intervene on outcome or selection variables.}
    \label{fig:SWIG}
\end{figure}

Figure \ref{fig:SWIG} illustrates three scenarios for measurement error and selection mechanisms using single world intervention graphs (SWIGs) \citep{richardson2013single}. 
As we discuss in our presentation of causal assumptions, SWIG (A) includes the simplest causal structure (i.e., strongest assumptions), enabling non-parametric identification of the ATE using gold-standard outcomes, selection variables, and observed covariates. SWIG (B) adds an arrow from one's gold-standard outcome to their selection indicator, complicating identification of the ATE, and SWIG (C) further incorporates causal relationships regarding silver-standard outcomes. SWIG (C) illustrates the causal relationships that we expect in ASPIRE, informing our identification assumptions as implied by the potential outcome calculus rules \citep{malinsky2019potential,shpitser2022multivariate}.

A key feature of SWIG (C) and its antecedent SWIG (B) is that treatment (i.e., the ``world"), covariates, and gold-standard outcomes all influence selection. Moreover, we allow treatment, in addition to covariates, to affect not only gold-standard outcomes but also silver-standard outcomes, inducing a form of differential measurement error \citep{carroll2006measurement,hernan2009invited}. We believe that homogeneity of (mis)classification with respect to treatment, as proposed by \cite{shu2019causal} and \cite{shen2025integrating}, is potentially an overly restrictive assumption in practice. For instance, in ASPIRE, whether clinics are assigned nudge or nudge+ may influence what their clinicians document in the EHR, and the two treatments may also result in different rates of parent-perceived program delivery. We also consider heterogeneity in rates of measurement error arising from covariates (e.g. patient, clinician, and site characteristics), which is likely present in visit-based studies. Simulation studies in Section \ref{sec:simstudy} demonstrate the danger of failing to account for this heterogeneity by treatment and possibly covariates.

The following identification assumptions, IA1-IA4, can be used to express the ATE in terms of observable data. 

\begin{itemize}

\item[\textbf{IA1}] Cluster-level SUTVA: Under $A_i=a \in \{0,1\}$, a) $Y_{ij}^*=Y^*_{ij}(a)$; b) $V_{ij}=V_{ij}(a)$; and c) $Y_{ij}=Y_{ij}(a)$ when $V_{ij}(a)=1$.

\item[\textbf{IA2}] Randomization: For $a=0,1$, $\{Y^*_i(a),Y_i(a),V_i(a),L_i,U_i\} \perp A_i$ where $0<P(A_i=1)<1$.

The terms $Y^*_i(a)=(Y^*_{ij}(a))_{j=1}^{n_i}$, $Y_i(a)=(Y_{ij}(a))_{j=1}^{n_i}$, and $V_i(a)=(V_{ij}(a))_{j=1}^{n_i}$, denote vectors of gold-standard, silver-standard, and selection potential outcomes, respectively, in cluster $i$ for treatment $a$. $L_i$ and $U_i$ represent bundles of all observed and unobserved covariates for cluster $i$, respectively, which can be at the individual- or cluster-level. 

\item[\textbf{IA3}] Classification ignorability: For $a=0,1$ and all individuals, $Y^*_{ij}(a) \perp V_{ij}(a)|Y_{ij}(a),X_{ij}$.

$X_{ij}$ is a subset of covariates $L_i$, defined for individual $j$ in cluster $i$, as illustrated in the SWIGs. Cluster-level characteristics are such that $X_{ij,p}=X_{ij',p}$ for all $j \ne j'$ in cluster $i$. These characteristics may describe the clusters directly, such as urban or suburban, or represent summary measures, such as cluster-average income \citep{raudenbush1997statistical}. 

\item [\textbf{IA4}] Differential classification: For $a=0,1$ and all individuals, 
\begin{equation}
P(Y^*_{ij}(a)=1|V_{ij}(a)=1,Y_{ij}(a)=1,X_{ij}) \ne P(Y^*_{ij}(a)=1|V_{ij}(a)=1,Y_{ij}(a)=0,X_{ij})
.\end{equation}
\end{itemize}

For the SUTVA assumption, IA1, we are assuming that there is one version of either treatment assigned to each cluster and there is no interference at the individual level. We are also making the implicit assumption that one's individual assigned treatment, $A_{ij}$, is such that $A_{ij}=A_i$, i.e., an individual's cluster assignment is not different from their own treatment assignment. Note that while individuals' gold-standard outcomes are only measurable if they are in the validation subset, their gold-standard potential outcomes are defined under both treatments in this context. This is in contrast to a setting where a terminal event is requisite for gold-standard ascertainment (e.g., for a brain autopsy for dementia \citep{suemoto2023autopsy}) since death could possibly preclude or obscure the measurement of one or more of the potential outcomes, a phenomenon of considerable interest in the causal inference literature referred to as ``truncation by death" \citep{zhang2003estimation}.

Assumptions IA1 and IA2 must be afforded to us by the design of the cluster-randomized trial so that we can target the individual-level ATE. One key observation is that under IA1 and IA2, we have internal validity \citep{mathur2025simple} for conditional classification probabilities, 
which signifies that potential outcome-based classification probabilities (LHS) are equated with observed classification probabilities (RHS) as follows: 
\begin{align}
\begin{split}
\label{eq:measprob}
P(Y^*_{ij}(a)=1|V_{ij}(a)=1,Y_{ij}(a)=y,X_{ij})&=P(Y_{ij}^*=1|V_{ij}=1,Y_{ij}=y,A_{ij}=a,X_{ij})\\
&:=p_{ij}^v(y,a,X_{ij}),
\end{split}
\end{align}
since IA2 is defined for all variables in  cluster $i$ \textit{jointly}. We note that this identity would hold even if there were an arrow from $X_{ij}$ to $A_{ij}$ in SWIG (C), that is if observed $X_{ij}$ were (also) sufficient to control for confounding, as in an observational setting, because it would confer conditional independence of potential outcomes from assignment given observed covariates. We briefly discuss this extension in Section \ref{sec:discussion}.

IA3 is necessary for targeting an estimand external to the validation subset. To clarify the role of IA3, we first note that IA3 is implied by the stronger assumption that for $a=0,1$, $\{Y_{ij}(a),Y^*_{ij}(a)\} \perp V_{ij}(a)|X_{ij}$, which we will denote IA3*. Unlike IA3, IA3* implies the selection ignorability condition $Y_{ij}(a) \perp V_{ij}(a)|X_{ij}$ as in SWIG (A). Therefore, under IA3*, we are not only able to identify the ATE through IA3 using silver-standard outcomes--correcting for misclassification as per our proposed approach--but also through the selection ignorability condition using gold-standard outcomes by accounting for non-random selection. The previous literature employed a version of IA3* to derive a most efficient estimator that is the optimal convex combination of the estimators associated with these two identified terms \citep{shen2025integrating,shu2019causal}.

To demonstrate why having the selection ignorability condition simplifies matters, we employ the selection mechanism framework of \cite{mathur2025simple}. They describe selection bias with respect to two components, ``internal bias" and ``net-external bias" conditional on $X_{ij}$. Their work extends \cite{lu2022toward}'s notions of Type 1 and Type 2 biases, respectively, allowing for treatment to affect selection. Consider
\begin{align}
\tau_{C}(X_{ij})&=E\{Y_{ij}(1)|X_{ij}\}-E\{Y_{ij}(0)|X_{ij}\}\\
\tau_{CV}(X_{ij})&=E\{Y_{ij}(1)|V_{ij}(1)=1,X_{ij}\}-E\{Y_{ij}(0)|V_{ij}(0)=1,X_{ij}\}\\
\tau_{\text{assoc}}(X_{ij})&=E\{Y_{ij}|V_{ij}=1,A_{ij}=1,X_{ij}\}-E\{Y_{ij}|V_{ij}=1,A_{ij}=0,X_{ij}\}
\end{align}
where conditional on $X_{ij}$ 
\begin{align}
\text{Internal Bias}&:=\tau_{\text{assoc}}(X_{ij})-\tau_{CV}(X_{ij})\\
\text{Net-external Bias}&:= \tau_{CV}(X_{ij})-\tau_{C}(X_{ij})\\
\text{Total Bias}&:= \tau_{\text{assoc}}(X_{ij})-\tau_C(X_{ij}).
\end{align}
Therefore, we can decompose the total (additive) bias, induced by selection, of the associational quantity relative to the conditional ATE as the sum of internal and net-external biases. Non-parametric identification (or V-identification in the parlance of Mathur and Shpitser) of the ATE, $E_{X_{ij}}\{\tau_C(X_{ij})\}$, with $E_{X_{ij}}\{\tau_{\text{assoc}}(X_{ij})\},$ would then require removing these two sources of bias.
Mathur and Shipster provide sufficient conditions for SWIGs (and their associated identification) conditional on $X_{ij}$ that would eliminate these sources of bias. Under cluster-level SUTVA (IA1) and the assumptions underpinning SWIG (A) (IA2 and selection ignorability), we would have those sufficient conditions for non-parametric identification of the ATE. In particular,  the ATE is identified with the typical standardization estimand or inverse probability of selection weighting (IPSW) estimand. As detailed in the Supplement Section \ref{sec:identification}, IA2 removes internal bias similar to Equation (\ref{eq:measprob}), 
and selection ignorability enables the key equality $E(Y_{ij}(a)|X_{ij})=E(Y_{ij}(a)|V_{ij}(a)=1,X_{ij})$ which would remove net-external bias. The IPSW identified estimand, $\tau_{\text{IPSW}}=\mu_{\text{IPSW}}(1)-\mu_{\text{IPSW}}(0)$, is
\begin{equation}
\label{eq:ipsw}
\mu_{\text{IPSW}}(1)-\mu_{\text{IPSW}}(0)=E\left\{\frac{A_{ij}V_{ij}Y_{ij}}{P(V_{ij}=1|A_{ij}=1,X_{ij})\pi}\right\}-E\left\{\frac{(1-A_{ij})V_{ij}Y_{ij}}{P(V_{ij}=1|A_{ij}=0,X_{ij})(1-\pi)}\right\},
\end{equation}
where $\pi=P(A_{ij}=1)$ and $V_{ij}Y_{ij}=0$ if $V_{ij}=0$\footnote{For estimation, we restrict to the subset of selected individuals instead of using this product notation such that the estimator is directly implementable in \texttt{R}.}. 

However, we consider the possibility that there is an arrow from $Y_{ij}(a)$ to $V_{ij}(a)$ as per SWIG (B) -- a relationship that, as discussed previously, is relevant to ASPIRE. Parents' perception of whether they received counseling and a cable lock at their child's visit may have influenced their willingness to respond to a follow-up survey. That is, outcome \textit{occurrence} preceded and thus possibly affected the selection mechanism (i.e., survey response), whereas outcome \textit{ascertainment} came after selection. Therefore, we consider that selection ignorability may not hold, and net-external bias would potentially persist without access to a more complete picture of variable relations than those provided in SWIG (B) or without resorting to parametric assumptions. In Section \ref{sec:simstudy}, we demonstrate the small-sample bias in the corresponding IPSW estimator (as compared to our proposed estimator) for simulated data where one's gold-standard outcome is used to generate their selection indicator (Figure \ref{fig:compareestim}).

SWIG (C) and the associated key condition IA3 provides us precisely with a more complete picture that fits our setting and still enables non-parametric identification. Namely, we leverage the relationship between the gold- and silver-standard outcomes, and the knowledge that one's silver-standard outcome does not impact their selection directly. As a necessary companion to IA3, IA4 requires that for every individual under both treatments, the conditional probability of a false positive misclassification is not the same as the probability of a true positive classification, which generally holds in practice, unless classification is deterministic. 

\begin{identity}
\label{identity1}
Under IA1-IA4, the ATE is identified with the silver-standard weighting (SSW) representation $\tau_{\text{SSW}}=\mu_{\text{SSW}}(1)-\mu_{\text{SSW}}(0)$ where
\begin{equation}
\label{eq:ssw}
\mu_{\text{SSW}}(1)-\mu_{\text{SSW}}(0)=E\left\{\frac{A_{ij}Y_{ij}^*-\pi p^v_{ij}(0,1,X_{ij})}{\pi[p^v_{ij}(1,1,X_{ij})-p^v_{ij}(0,1,X_{ij})]}\right\}-E\left\{\frac{(1-A_{ij})Y_{ij}^*-(1-\pi) p^v_{ij}(0,0,X_{ij})}{(1-\pi)[p^v_{ij}(1,0,X_{ij})-p^v_{ij}(0,0,X_{ij})]}\right\}.
\end{equation}
\end{identity}
We briefly outline the proof of Proposition \ref{identity1} below with full details provided in Supplement Section \ref{sec:identification}. Applying IA3 and Equation (\ref{eq:measprob}) (via IA1 and IA2) affords us net-external validity regarding the conditional classification probabilities 
\begin{align}
\begin{split}
\label{eq:extvern}
P(Y^*_{ij}(a)=1|Y_{ij}(a)=y,X_{ij})&\stackrel{\text{IA3}}{=}P(Y^*_{ij}(a)=1|V_{ij}(a)=1,Y_{ij}(a)=y,X_{ij})\\
&\stackrel{\text{Eq\ref{eq:measprob}}}{=}P(Y^*_{ij}=1|V_{ij}=1,Y_{ij}=y,A_{ij}=a,X_{ij})=p_{ij}^v(y,a,X_{ij}).
\end{split}
\end{align}
Therefore, the terms $P(Y^*_{ij}(a)=1|Y_{ij}(a)=y,X_{ij})$ are estimable using observed data from the selection subset. Looking ahead, depending on the structure of $X_{ij}$ and its relationship to $Y^{*}_{ij}$ in these probabilities, estimation may require modeling assumptions, albeit as flexibly as desired to avoid mis-specification. The essential piece for determining the identity in Equation (\ref{eq:ssw}) is that $E(Y^*_{ij}(a)|X_{ij})$ can be written in two ways for $a=0,1$. By IA1 and IA2, we have
\begin{equation}
\label{eq:eysLHS}
E(Y^*_{ij}(a)|X_{ij})\stackrel{\text{IA2}}{=}E(Y^*_{ij}(a)|A_{ij}=a,X_{ij})\stackrel{\text{IA1-2}}{=}E(Y_{ij}^*I(A_{ij}=a)|X_{ij})\left[P(A_{ij}=a)\right]^{-1}.
\end{equation}
Using (\ref{eq:extvern}) and the law of total expectation (LOTE), we can also write
\begin{align}
\label{eq:eysRHS}
\begin{split}
E(Y^*_{ij}(a)|X_{ij}) \stackrel{\text{IA1-4}}{=}p_{ij}^v(1,a,X_{ij})E(Y_{ij}(a)|X_{ij})+p_{ij}^v(0,a,X_{ij})\left[1-E(Y_{ij}(a)|X_{ij})\right].
\end{split}
\end{align}
Setting Equations (\ref{eq:eysLHS}) and  (\ref{eq:eysRHS}) equal, we solve for $E(Y_{ij}(a)|X_{ij})$. Finally, we use $E\{E(Y_{ij}(a)|X_{ij})\}=E\{Y_{ij}(a)\}$ (LOTE) to obtain the estimable quantities $\mu_{\text{SSW}}(0)$ and $\mu_{\text{SSW}}(1)$. IA4 ensures that the term $E(Y_{ij}(a)|X_{ij})$ remains in Equation (\ref{eq:eysRHS}), preventing division by zero. 
 
\section{Estimation of the ATE}
\label{sec:estimation}

We provide a plug-in estimator for the ATE, where we take a difference of empirical means and plug in the requisite estimated classification and treatment probabilities, following the expression for $\tau_{\text{SSW}}$ (\ref{eq:ssw}). For asymptotic properties, we represent our estimator as a linear combination of the solution to cluster-indexed unbiased estimation equations (UEEs), constructed by stacking or combining \citep{stefanski2002calculus} the UEEs for the classification and treatment probabilities, and those for the $\mu_{\text{SSW}}(a)$ terms given those probabilities; we then appeal to M-estimation theory \citep{huber1964robust,huber1967}. In some settings, it may be that the special case of $X_{ij}=\emptyset$ for assumptions IA3 and IA4 is justified (i.e., the olive arrow in SWIG (C) is removed and no subset of $L_i$ is necessary for identification). This condition, which we will henceforth refer to as condition i), has considerable appeal since it implies $Y^*_{ij}$ need only be represented in terms of $Y_{ij}$ and $A_{ij}$, and the ATE would be estimable non-parametrically using empirical estimators for the component classification probabilities. If we cannot assume condition i), we may need to propose a model for the classification probabilities and necessarily so if $X_{ij}$ were continuous. To this end, we opt for generalized estimating equations (GEE) with a parametric mean model and variance function as per logistic regression and an independence working correlation \citep{liang1986longitudinal}. We opt for this model since it is enables cluster-robust variance expressions  and fits nicely within our M-estimation framework. Our development of the SSW estimator via UEEs below encapsulates both when condition i) does and does not hold.  

Importantly, the proposed SSW estimator for the ATE uses all the available data, including silver-standard outcomes as well as gold-standard outcomes from the subset of individual units for which $Y_{ij}$ is recorded (i.e., with $V_{ij}=1$). In simpler settings, under modest rates of measurement error, this estimator has been shown to boost efficiency compared to an IPSW estimator that uses only the correctly measured outcomes from the (much) smaller validation subset \citep{shu2019causal,shen2025integrating}. In addition, the above identification, which does not require the selection ignorability condition, suggests that the associated SSW estimator--unlike the IPSW estimator or an analogous standardization-based estimator--can handle the situation where an individual's gold-standard outcome influences their probability of being in the selection subset, i.e., the gold-standard outcomes are missing not at random \citep{daniel2012using,seaman2013review,li2013weighting}. In exchange, our estimator relies on modeling the classification mechanism instead of specifying an outcome model via the selection subset and/or a selection propensity score model. We investigate these ideas further with simulation studies in Section \ref{sec:simstudy}.

\subsection{Classification Modeling}

Recall that $\{Y^*_i,Y_i,V_i,L_i\}_{i=1}^m$ are the observed data in each cluster $i$, where we define $Y^*_i=(Y^*_{ij})_{j=1}^{n_i}$, $V_i=(V_{ij})_{j=1}^{n_i}$, and $Y_i=(Y_{ij})_{j:V_{ij}=1}$. We assume these data are independent with respect to $i$. Moreover, for a given $n_i$, $Y^*_i$ and $V_i$ are identically distributed across clusters, and for a given $n_i^v$, $Y_i$ is identically distributed across clusters. We established through identification that the probabilities $P(Y^*_{ij}(a)=1|Y_{ij}(a)=y,X_{ij})$ are estimable using $P(Y^*_{ij}=1|V_{ij}=1,Y_{ij}=y,A_{ij}=a,X_{ij})$ as modeled on the validation subset. We will assume that the classification model follows a logistic GEE with an independence working correlation. The parametric mean model is
 \begin{equation}
\label{eq:geemean}
P(Y^*_{ij}=1|V_{ij}=1,Y_{ij},A_{ij},X_{ij})=p^v_{ij}(D_{ij};\theta)=\text{expit}(D_{ij}^T\theta),
\end{equation}
where $D^T_{ij}=(1,Y_{ij},A_{ij},Y_{ij}A_{ij},W_{ij}^T)$ is a row of the design  matrix $D_i$ for cluster $i$. $W_{ij}$ includes any necessary remaining regressors from the components of $X_{ij}$ and their interactions. $\theta$ represents the regression coefficients. $D_{ij}$ is only completely defined in the observed data on the selection subset when the gold-standard $Y_{ij}$ is recorded, and thus
we restrict to the $D_{ij}$ where $V_{ij}=1$ for model fitting. While our model is defined generally, we will only take into consideration regressor interactions with $A_{ij}$. 

Let $p^v_i(D_i;\theta)$ be a column vector with components $p^v_{ij}(D_{ij};\theta)$ for cluster $i$. Cluster covariance matrices are defined as
\begin{equation}\Sigma^v_i =\text{diag}_{n^v_i}\left (p^v_{i}(D_i;\theta) \odot [1-p^v_{i}(D_i;\theta)]\right),
\end{equation}
where $\odot$ denotes the element wise or Hadamard product, and $\text{diag}_{n^v_i}$ puts the elements of the input vector on the diagonal of an $n^v_i \times n^v_i$ diagonal matrix. This specification sets up the following GEE among the selection subset whose solution produces estimates of regression parameters, $\hat \theta$.
\begin{equation}
\label{ref:gee_ee}
\sum_{i=1}^{m}\sum_{j:V_{ij}=1} D_{ij}\left(Y_{ij}^*-\frac{\exp(D_{ij}^T\theta)}{1+\exp(D_{ij}^T\theta)}\right)=0\end{equation}

We obtain the estimator, $\hat \tau_{\text{SSW}}$, as a linear transformation of the solution to stacked cluster-indexed UEEs with estimating functions $m_i(\lambda)$ whose dimension equals the parameter vector $\lambda^T=(\theta^T,\pi,\mu_{\text{SSW}}(1),\mu_{\text{SSW}}(0))$. Specifically, we stack the GEE ``on top of" the UEEs defined according to the terms $\pi$, $\mu_{\text{SSW}}(1)$, and $\mu_{\text{SSW}}(0)$ of Equation (\ref{eq:ssw}): 
\begin{equation}
\sum_{i=1}^m m_i(\lambda)=\sum_{i=1}^m
\begin{bmatrix}
\displaystyle\sum_{j:V_{ij}=1}D_{ij}\left(Y_{ij}^*-\frac{\exp(D_{ij}^T\theta)}{1+\exp(D_{ij}^T\theta)}\right)\\
 \displaystyle \sum_{j=1}^{n_i}(A_{ij}-\pi)\\
 \displaystyle \sum_{j=1}^{n_i} \left(\frac{A_{ij}Y^*_{ij}- \pi p^v_{ij}(D_{ij,01};\theta)}{ \pi [ p^v_{ij}(D_{ij,11};\theta)- p^v_{ij}(D_{ij,01};\theta)]}-\mu_{\text{SSW}}(1)\right)\\
\displaystyle \sum_{j=1}^{n_i}\left(\frac{(1-A_{ij})Y^*_{ij}-(1-\pi) p^v_{ij}(D_{ij,00};\theta)}{(1- \pi) [p^v_{ij}(D_{ij,10};\theta)- p^v_{ij}(D_{ij,00};\theta)]}-\mu_{\text{SSW}}(0)\right)
\end{bmatrix}=0.
\end{equation}
$D_{ij,ya}$ signifies setting $Y_{ij}=y \in \{0,1\}$ and $A_{ij}=a \in \{0,1\}$ for $D_{ij}$. The solution to these estimating equations is $\hat \lambda^T=(\hat \theta^T,\hat \pi, \hat \mu_{\text{SSW}}(1),\hat \mu_{\text{SSW}}(0))^T$. Our estimator is then $\hat \tau_{\text{SSW}}=k^T\hat\lambda$ for $k=(0^T,0,1,-1)^T$, which equals
\begin{equation} 
\label{eq:sswestimator}
\hat \tau_{\text{SSW}}=\frac{1}{N} \sum_{i=1}^m \sum_{j=1}^{n_i} \frac{A_{ij}Y^*_{ij}-\hat \pi \hat p^v_{ij}(D_{ij,01};\hat \theta)}{\hat \pi [\hat p^v_{ij}(D_{ij,11};\hat \theta)-\hat p^v_{ij}(D_{ij,01};\hat \theta)]}- \frac{1}{N}\sum_{i=1}^m \sum_{j=1}^{n_i}\frac{(1-A_{ij})Y^*_{ij}-(1-\hat \pi) \hat p^v_{ij}(D_{ij,00};\hat \theta)}{(1-\hat \pi) [\hat p^v_{ij}(D_{ij,10};\hat \theta)-\hat p^v_{ij}(D_{ij,00};\hat \theta)]} \end{equation}
with $\hat p^v_{ij}(D_{ij,ya};\hat \theta)=\expit(D_{ij,ya}^T\hat \theta)$. Under certain regularity conditions, a first order Taylor expansion per M-estimation theory \citep{huber1964robust,huber1967} and application of the delta method lead us to $\sqrt{m}(\hat \tau_{\text{SSW}}-\tau_0) \xrightarrow{d} N(0,k^TV_{\hat \lambda}k)$ \citep{liang1986longitudinal,stefanski2002calculus}. $\tau_0$ represents the true ATE, assuming correct specification of the classification model. $V_{\hat \lambda}$ denotes the true asymptotic variance of the estimator $\hat \lambda$, which has a sandwich form. We approximate $V_{\hat \lambda}$ using the sandwich estimator
\begin{equation}
\label{eq:crob}
\hat V_{\hat \lambda}=\left[\frac{1}{m}\sum_{i=1}^{m} \left \{\frac{\partial m_{i}(\hat \lambda)}{\partial \lambda^T}\right \} \right]^{-1} \frac{1}{m}\sum_{i=1}^{m}m_{i}(\hat \lambda)m_{i}(\hat \lambda)^T  \left[\frac{1}{m} \sum_{i=1}^{m} \left \{\frac{\partial m_{i}(\hat \lambda)}{\partial \lambda^T} \right \}\right]^{-T}, 
\end{equation}
where $\frac{\partial m_{i}(\hat \lambda)}{\partial \lambda^T}$ denotes the derivative (Jacobian) of $m_i$ with respect to $\lambda$ evaluated at $\hat \lambda$ \citep{iverson1989effects}. We provide this derivative in the Supplement Section \ref{sec:uee}. By indexing our UEEs by cluster, this sandwich variance expression accounts for cluster-level correlations (via the middle matrix) even if the modeling assumptions in the UEEs do not do so correctly or at all (via the outside matrices). Therefore, it should be robust to clustering effects and thus is often referred to as a cluster-robust variance estimator. Finally, we note that this M-estimation construction is applicable to independent and identically distributed (iid) data as well when the estimating equations are indexed with respect to each individual in lieu of cluster.

Under the simplifying condition i), we would have the special case that $W_{ij}$ is empty (since $X_{ij}$ is empty), and the model for (\ref{eq:geemean}) would only include $D^T_{ij}=(1,Y_{ij},A_{ij},Y_{ij}A_{ij})$. In this situation, the GEE would be saturated, and the terms $\hat p^v_{ij}(D_{ij,ya};\hat \theta)$ (by substitution) are in fact estimated non-parametrically as 
\begin{equation}
\label{eq:nonpar}
\hat p^v_{ij}(D_{ij,ya};\hat \theta)=\hat p^v(y,a)= \frac{\sum_{i=1}^m \sum_{j:V_{ij}=1} Y^*_{ij}I(Y_{ij}=y)I(A_{ij}=a)}{\sum_{i=1}^m \sum_{j:V_{ij}=1} I(Y_{ij}=y)I(A_{ij}=a)} .
\end{equation}
In other words, for consistent point estimation of the ATE under condition i), we do not need to directly specify any model to estimate the requisite marginal probabilities across treatments. This feature is attractive since we would avert the potential consequences of model mis-specification in estimating the ATE. However, if condition i) were not to hold, i.e., $W_{ij}$ were not empty, using the marginal estimates (\ref{eq:nonpar}) may not suffice to target the ATE consistently, and we would fall back on our parametric estimation. We thus explore the role of correctly or falsely assuming condition i) on the small-sample operating characteristics in simulations. 

Lastly, as with earlier approaches \citep{shu2019causal,shen2025integrating} for correcting misclassification error, particular estimates for $\hat \tau_\text{SSW}$ may not necessarily fall in $[-1,1]$ because $\hat \mu_\text{SSW}(1)$ and $\hat \mu_\text{SSW}(0)$ are not inherently constrained to $[0,1]$. These estimators are solutions to systems of equations where we have used plug-in estimators for the various nuisance parameters involved. Hence, not unlike Horvitz-Thompson inverse probability of treatment weighting (IPTW) estimation, there is no guarantee our estimators respect all the bounds of the true parameters and estimands in small samples \citep{rotnitzky2012improved}. In practice, we would expect this issue to arise when the true proportions $\mu(1)$ and $\mu(0)$ are near to 0 or 1 and when the sample size is not sufficient to distinguish small deviations from 0 or 1. However, while the estimated proportions may be slightly outside of 0 to 1, the ATE (i.e., risk difference) may still result in a value that respects its bounds as shown in simulations. 

\section{Simulation Study}
\label{sec:simstudy}

 We first consider the small-sample performance of our proposed estimator and its  cluster-robust variance estimator (Table \ref{tab:sim_table1}). Let $\mathbb{I}_d$ and $\mathbb{J}_d$ represent the $d \times d$ identity matrix and matrix of all 1's respectively. Denote uniform with $\mathcal{U}$, where brackets refer to the discrete distribution and parentheses to the continuous. We generate 5,000 simulated data sets according to 8 factorial scenarios given by: number of clusters $m=30$, cluster sizes $N_i \sim \mathcal{U}\{100,300\}$ or $N_i \sim \mathcal{U}\{500,1000\}$, intracluster correlation coefficient (ICC) values of 0.01 or 0.1, and whether or not the true classification model among potential outcomes includes covariates. For each cluster, we have $X_{i1} \sim \MVN(\mathbf{1},\mathbb{I}_{n_i})$, $X_{i2} \sim \MVN(\mathbf{.5},.5\mathbb{I}_{n_i}+.05\mathbb{J}_{n_i})$, $X_{i3}$ such that $X_{ij3} \overset{\text{iid}}{\sim} \Bern(.55)$, and cluster-level $X_{i4} \sim \mathcal{U}(0,1)$. We generate the data according to the following general models:
\begin{align}
\label{eq:dgm}
    & Y_{ij}(a) \sim \Bern\left\{\expit(\psi_{ij}(a,X_{ij},b_{i,y}) \right\} & &\psi_{ij}(a,X_{ij},b_{i,y})=\alpha_{a,y}+\beta_{a,y}^TX_{ij}+b_{i,y} \nonumber\\
    & Y^*_{ij}(a) \sim \Bern\left\{\expit(\rho_{ij}(a,Y_{ij}(a),X_{ij}) \right\} & &\rho_{ij}(a,Y_{ij}(a),X_{ij}) = \alpha_{a,y^*}+\beta_{a,y^*}^{T}X_{ij}+\delta_{a,y^*}Y_{ij}(a)\\
    & V_{ij}(a) \sim \Bern\left\{\expit(\varphi_{ij}(a,Y_{ij}(a),X_{ij},b_{i,v}) \right\} & & \varphi_{ij}(a,Y_{ij}(a),X_{ij},b_{i,v}) = \alpha_{a,v}+\beta_{a,v}^{T}X_{ij}+\delta_{a,v}Y_{ij}(a)+b_{i,v}, \nonumber 
\end{align}
where $b_{i,y} \overset{\text{iid}}{\sim} N(0,\sigma^2_{b,y})$ and $b_{i,v} \overset{\text{iid}}{\sim} N(0,\sigma^{2}_{b,v})$, which are independent of all regressors. We set these variances such that the ICC values, defined on the response scale by $\text{ICC}=\sigma_{b,(\cdot)}^2/(\sigma^2_{b,(\cdot)}+\pi^2/3)$ \citep{eldridge2009intra}, are either 0.01 or 0.1. Observed data in each iteration are generated according to the assignment mechanism $A_i \sim \Bern(.5)$, where the simulated potential outcomes become observed under their respective assignment. For example, if $A_i=1$, the observed data are the $V_{ij}=V_{ij}(1)$, $Y^*_{ij}=Y^*_{ij}(1)$, and $Y_{ij}=Y_{ij}(1)$ if $V_{ij}(1)=1$ or is NA if $V_{ij}(1)=0$ in cluster $i$. The true value of the ATE estimand is defined as the differences in means, $\overline{Y_{ij}(1)}-\overline{Y_{ij}(0)}$, averaged across 5,000 iterations. 

For Table \ref{tab:sim_table1}, the model parameters are 
\begin{align}
\label{eq:paramst1}
\begin{split}
(\alpha_{a,y},\beta^T_{a,y})&=(-1+0.75a,0.15,0.2,0.15,-0.15) \\
(\alpha_{a,y^*},\beta^{T}_{a,y^*},\delta_{a,y^*})&=  \begin{cases}
    (-1.25+0.25a,0^T,1.5+a) & \text{NDX}\\
    (-1.25+0.5a,0.25-0.5a,-0.25+0.1a,-0.15-0.1a,0.1,1.5+a)   & \text{DX}
  \end{cases}\\
(\alpha_{a,v},\beta^T_{a,v},\delta_{a,v})&=(-0.25-0.25a,-0.5,-0.5,0.25,-0.25,-0.15+0.3a)
\end{split}
\end{align}
where DX and NDX denote whether generation of $Y^*_{ij}(a)$ depends on covariates or does not depend on covariates (as per condition i)), respectively. As stated earlier, regardless of the role of covariates, we allow for classification probabilities to depend on treatment. The parameters were chosen to target an observed selection probability of 25\%-30\% and total classification error of 25\%-30\% among those selected. Note, these target proportions, along with all remaining ones in this section, were computed as the average proportions of 5,000 simulations in the observed data. The type of measurement errors for the observed $(Y_{ij},Y^*_{ij})$ are $(1,0)$ and $(0,1)$. Across the various data-generating scenarios, the average proportion of $(1,0)$ errors range from 9\%-10\% under treatment 1 and 12\%-13\% under treatment 0, and $(0,1)$ errors occurred at rate of about 13\% under treatment 1 and 16\%-17\% for treatment 0. In the context of ASPIRE, these values would indicate that clinicians reported delivering SAFE Firearm when parents believed that they had not received it more frequently than the reverse and that both misclassification types were increased under the nudge as opposed to the nudge+ arm.   

In supplemental simulations of Supplement Section \ref{sec:addtables}, we consider a hierarchical structure of clinician nested within cluster (i.e., site) to demonstrate robustness of our cluster-indexed M-estimator when there are just a handful of clinicians per site on average (as with ASPIRE). We set the within-site within-clinician correlation to 0.01 or 0.1 and make it double the within-site across-clinician correlation \citep{arnup2017understanding} such that the additional variation in outcomes due to clinician is equal to that of the site itself as described in the Supplement. For the data-generating mechanism per Table \ref{tab:sim_table1}, we have assumed that there are no random intercepts $b_{i,y^*}$ in $\rho_{ij}(a,Y_{ij}(a),X_{ij})$ so that the marginal model $p^v_{ij}(D_{ij};\theta)$ would not be inherently misspecified; we note that the conditional independence statement i) does not account for non-trivial $b_{i,y^*}$ either. In Supplement Section \ref{sec:addtables}, we also examine sensitivity to this assumption, and the results show that our estimator performs satisfactorily under realistic levels of ICC induced by $b_{i,y^*} \overset{\text{iid}}{\sim} N(0,\sigma^2_{b,y^*})$.

\begin{table}[H]
    \centering
    \captionsetup{labelfont=bf,justification=raggedright,singlelinecheck=false}
    
    \caption{Evaluation of the small-sample performance of our proposed estimation method under differential misclassification with a non-random internal validation set. We generated data for 30 clusters as described in Section \ref{sec:simstudy} with reference Equations (\ref{eq:dgm}) and (\ref{eq:paramst1}). 8 scenarios are reported which have two levels of ICC=0.01 or 0.1, two levels of cluster sizes, $\mathcal{U}\{100,300\}$ or $\mathcal{U}\{500,100\}$, and exclusion or inclusion of covariates in the classification model generating silver-standard outcomes. Empirical bias and coverage of approximate  95\% normal intervals and small-sample corrected $t$-intervals using the cluster-robust (CR) sandwich variance are reported. The model-based CR variances are compared to the empirical variances for reference. Model 1 which accounts for covariates in fitting a classification model is juxtaposed against Model 2 which has no covariates, i.e., Model 2 estimates classification probabilities non-parametrically.}
    
   \begin{tabular}{>{\centering\arraybackslash}p{1.5cm} 
                >{\centering\arraybackslash}p{1.5cm} 
                >{\centering\arraybackslash}p{1.5cm} 
                >{\centering\arraybackslash}p{1.5cm} 
                >{\centering\arraybackslash}p{1.5cm} 
                >{\centering\arraybackslash}p{1.5cm} 
                >{\centering\arraybackslash}p{1.5cm}}
    \toprule
    \textbf{\thead{Model}} & \textbf{\thead{True \\ ATE}} & \textbf{\thead{Empirical \\ Bias}} & \textbf{\thead{Empirical \\ Variance}} & \textbf{\thead{C-Robust \\ Variance}} & \thead{\textbf{Coverage}} & \textbf{\thead{Corrected \\ Coverage}} \\
    \midrule
    \multicolumn{7}{c}{\textit{True Classification Model Excludes Covariates}} \\
    \addlinespace
    \multicolumn{7}{l}{ICCs: 0.010; Cluster Sizes: $\mathcal{U}\{100,300\}$}\\
    \midrule
    Model 1 & 0.176 & -0.003 & 0.004 & 0.003 & 93.2\% & 94.3\% \\
    Model 2 & 0.176 & -0.001 & 0.003 & 0.003 & 93.1\% & 94.2\% \\
    \addlinespace
    \multicolumn{7}{l}{ICCs: 0.010; Cluster Sizes: $\mathcal{U}\{500,1000\}$} \\
    \midrule
    Model 1 & 0.176 & -0.001 & 0.001 & 0.001 & 92.2\% & 93.6\% \\
    Model 2 & 0.176 & -0.000 & 0.001 & 0.001 & 92.2\% & 93.7\% \\
    \addlinespace
    \multicolumn{7}{l}{ICCs: 0.100; Cluster Sizes: $\mathcal{U}\{100,300\}$}\\
    \midrule
    Model 1 & 0.165 & -0.002 & 0.006 & 0.005 & 92.4\% & 93.9\% \\
    Model 2 & 0.165 & -0.001 & 0.005 & 0.005 & 93.1\% & 94.4\% \\
    \addlinespace
    \multicolumn{7}{l}{ICCs: 0.100; Cluster Sizes: $\mathcal{U}\{500,1000\}$} \\
    \midrule
    Model 1 & 0.165 & -0.001 & 0.003 & 0.003 & 92.7\% & 94.2\% \\
    Model 2 & 0.165 & -0.000 & 0.003 & 0.003 & 92.7\% & 94.1\% \\
    \bottomrule
    \addlinespace
    \multicolumn{7}{c}{\textit{True Classification Model Includes Covariates}} \\
    \addlinespace
    \multicolumn{7}{l}{ICCs: 0.010; Cluster Sizes: $\mathcal{U}\{100,300\}$}\\
    \midrule
    Model 1 & 0.176 & -0.002 & 0.004 & 0.003 & 93.0\% & 94.2\% \\
    Model 2 & 0.176 & -0.062 & 0.003 & 0.003 & 76.1\% & 79.4\% \\
    \addlinespace
    \multicolumn{7}{l}{ICCs: 0.010; Cluster Sizes: $\mathcal{U}\{500,1000\}$} \\
    \midrule
    Model 1 & 0.176 & -0.001 & 0.001 & 0.001 & 92.3\% & 94.0\% \\
    Model 2 & 0.176 & -0.060 & 0.001 & 0.001 & 45.9\% & 50.4\% \\
    \addlinespace
    \multicolumn{7}{l}{ICCs: 0.100; Cluster Sizes: $\mathcal{U}\{100,300\}$} \\
    \midrule
    Model 1 & 0.165 & -0.001 & 0.006 & 0.005 & 92.5\% & 94.1\% \\
    Model 2 & 0.165 & -0.058 & 0.005 & 0.005 & 84.9\% & 87.2\% \\
    \addlinespace
    \multicolumn{7}{l}{ICCs: 0.100; Cluster Sizes: $\mathcal{U}\{500,1000\}$} \\
    \midrule
    Model 1 & 0.165 & -0.000 & 0.003 & 0.003 & 92.3\% & 93.9\% \\
    Model 2 & 0.165 & -0.056 & 0.003 & 0.003 & 80.4\% & 83.1\% \\
    \bottomrule
\end{tabular}
    \label{tab:sim_table1}
    \caption*{Model 1 - includes covariates in fitted classification model. Model 2 - does not include covariates in fitted classification model. Corrected coverage - coverage of $t$-intervals with $m-7$ degrees of freedom. ICC - intracluster correlation coefficient.}
\end{table}

To demonstrate performance, Table \ref{tab:sim_table1} includes the empirical bias of our estimator and the coverage of approximate 95\% normal confidence intervals computed using the cluster-robust sandwich variance estimator. This variance estimator is known to underestimate the true variance in small samples of clusters \citep{kahan2016increased,li2015small}.
Therefore, we also consider a $t$-interval with $m-7$ degrees of freedom (as there are seven estimated probabilities in defining $\tau$) to correct for potential undercoverage \citep{li2015small}. Lastly, we include a column comparing the sample variance of our estimator over the simulations, referred to as empirical variance, to the average estimated cluster-robust variance; we would expect these to be similar if the model-based variance is appropriate. 

Model 1 denotes the estimator for which the logistic GEE  classification model includes covariates $X_{ij}$ in its mean expression and is correctly specified according to DX. Specifically, we would have $D^T_{ij}=(1,Y_{ij},A_{ij},X_{ij,1:3}^TA_{ij},X_{ij,4})$ in the mean model (\ref{eq:geemean}), where 1:3 indicates components 1 through 3. Model 2 drops the covariates (assuming condition i)) such that all measurement error probabilities can be estimated non-parametrically via Equation (\ref{eq:nonpar}). We see that across all scenarios, whether or not the data generation of silver-standard outcomes is independent of covariates, Model 1 achieves low bias and has near nominal coverage, particularly after small-sample corrections. When classification is independent of covariates as per NDX, Model 2 performs satisfactorily as well, and the empirical bias and variance tend to be less than the over-parameterized Model 1 estimator. However, if we had errantly estimated the measurement probabilities non-parametrically when they truly depend on covariates, we see considerable bias and poor coverage across all scenarios. Therefore, if we cannot definitively rule out the role of covariates in classification probabilities, Model 1 may be the safest choice, acknowledging the risk of sacrificing efficiency in exchange for ensuring good coverage and low bias. 

We found some minor undercoverage of the estimators across all scenarios when the classification model is not mis-specified, suggesting a more aggressive small-sample correction might be warranted. However, in Table \ref{tab:sim_tableapp1} of Supplement Section \ref{sec:addtables}, we show that the $t$-interval using the asymptotic variance estimator achieves coverage comparable to the non-parametric cluster bootstrap \citep{field2007bootstrapping} (with 100 to 300 individuals per cluster and 1000 bootstrap iterations), where the latter is significantly more computationally expensive. Moreover, both of these cluster-based methods perform better than their individual counterparts (treating the data as iid), which result in significant undercoverage when ICCs and cluster sizes are larger. In the remaining supplemental tables, we examine the robustness of the SSW estimator and the associated confidence interval that do not account for some element of the hierarchical structure. Our estimator displayed low bias both when we added a clinician random intercept in outcome and selection generation (Table \ref{tab:sim_tableapp2}) and to a cluster random intercept in the classification model (Table \ref{tab:sim_tableapp3}). Similar coverage patterns are observed in these supplemental tables as in the main data-generating mechanism, although undercoverage appears slightly worse when clinician-level random intercepts are included, as anticipated.

\begin{figure}[H]
  \centering
  \captionsetup{labelfont=bf,justification=raggedright,singlelinecheck=false}
  \includegraphics[width=1\linewidth,scale = 1]{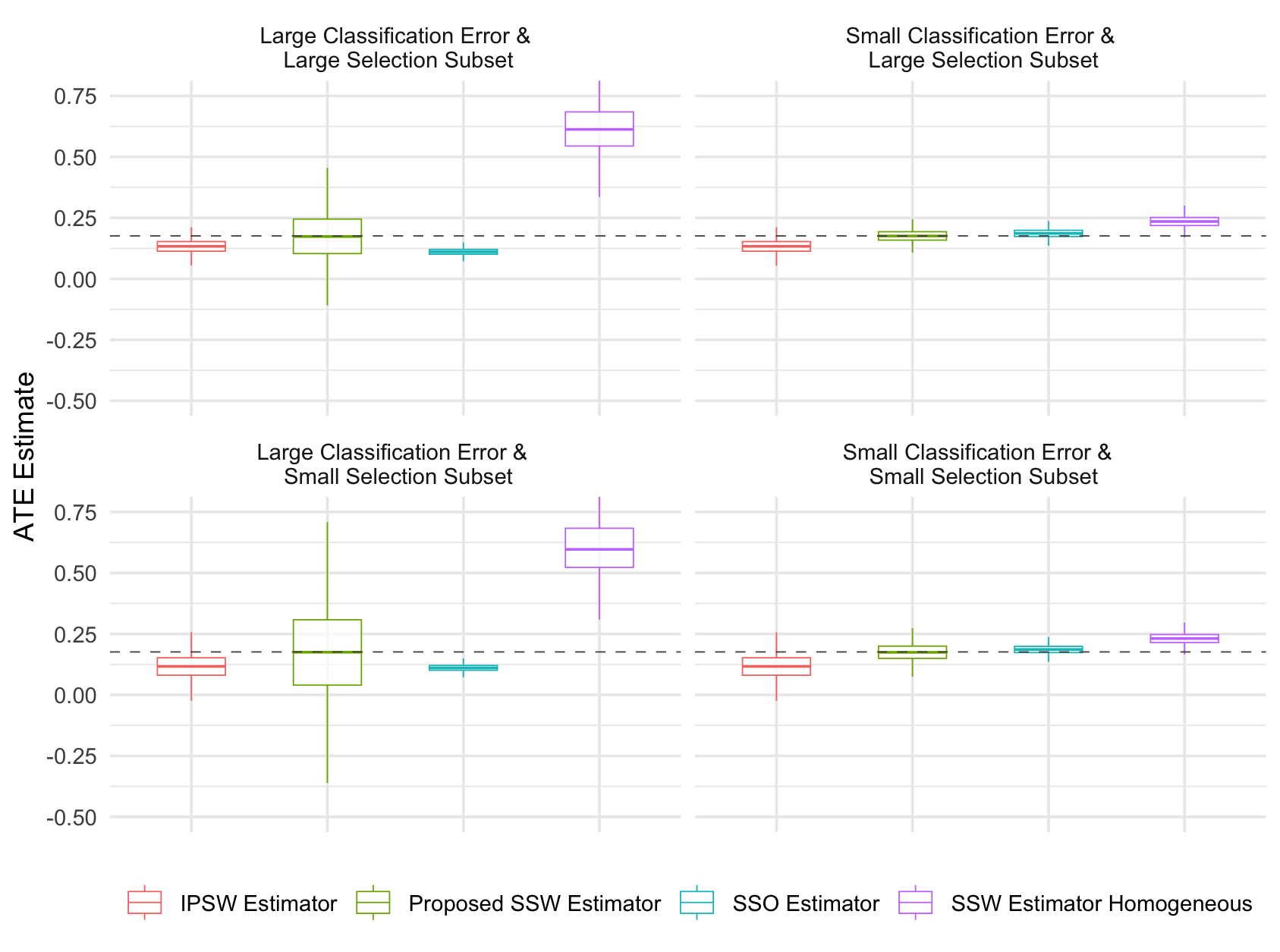}
  \centering
  \caption{Comparing estimators across four different combinations of selection subset proportions and classification error rates. Observed large selection proportion is just over $40\%$ (about an average of 2400 individuals) and small selection proportion is just under $20\%$ (about an average of 1200 individuals). Small misclassification is approximately $9\%$-$10\%$ and large is approximately $40\%$-$41\%$; rate and/or direction of misclassification vary by treatment. IPSW estimator stands for inverse probability of selection weighting estimator, $\hat \tau_{\text{IPSW}}$. SSW estimator represents our proposed silver-standard weighting estimator, $\hat \tau_{\text{SSW}}$. SSO estimator denotes the silver-standard only estimator, $\hat \tau_{\text{SSO}}$. Homogeneous denotes that the fitted classification model for relationship between silver- and gold-standard outcomes does not depend on treatment or covariates.}

  \label{fig:compareestim}
\end{figure}

In Figure \ref{fig:compareestim}, we compare the small-sample bias and variability of four estimators under different rates of measurement error and sizes of the validation subsets. The four estimators shown are: a) our proposed SSW estimator $\hat \tau_{\text{SSW}}$ with correct classification model specification (Model 1) in green, b) the gold-standard IPSW estimator $\hat \tau_{\text{IPSW}}$ in red (defined below), c) the silver-standard only (SSO) estimator  $\hat \tau_{\text{SSO}}$ in blue (defined below), and d) the SSW estimator but with a classification model that is homogeneous across treatment and covariates \citep{shen2025integrating,shu2019causal} in purple. The SSO estimator targets the error-prone (e.g. clinician-reported) ATE, and its expression is simply the difference in means, which we write as
\begin{equation}
\label{eq:sso_estim}
\hat \tau_{\text{SSO}}=\frac{1}{N} \sum_{i=1}^m \sum_{j=1}^{n_i} \frac{A_{ij}Y^*_{ij}}{\hat \pi}-\frac{1}{N} \sum_{i=1}^m \sum_{j=1}^{n_i} \frac{(1-A_{ij})Y^*_{ij}}{(1-\hat \pi)} \approx E\{Y^*_{ij}(1)\}-E\{Y^*_{ij}(0)\},
\end{equation}
where $\hat \pi$ is the empirical proportion of treated individuals. In general, the SSO estimator should not target the ATE of interest without further assumptions. For example, \cite{shu2019causal} demonstrate consistency of this estimator under correct specification of a  measurement error model that has an identity link mean function and specific linear representations in its treatment and gold-standard outcome arguments. The plug-in IPSW estimator, which is defined within the validation subset, is
\begin{equation}
\label{eq:ipsw_estim}
\hat \tau_{\text{IPSW}}=\frac{1}{N}\sum_{i=1}^m \sum_{j:V_{ij}=1}\frac{A_{ij}Y_{ij}}{\hat P(V_{ij}=1|A_{ij}=1,X_{ij})\hat \pi}-\frac{1}{N}\sum_{i=1}^m \sum_{j:V_{ij}=1}\frac{(1-A_{ij})Y_{ij}}{\hat P(V_{ij}=1|A_{ij}=0,X_{ij})(1-\hat \pi)},
\end{equation}
which targets the IPSW representation in Equation (\ref{eq:ipsw}), but it may not target the ATE unbiasedly when $V_{ij}(a)$ and $Y_{ij}(a)$ are related as per SWIG (B).

We define a low and high proportion of selected individuals as approximately 20\% and 40\%, respectively. Among those selected, we define overall low and high measurement error (i.e, $(Y_{ij},Y^*_{ij})=(1,0)$ or $(0,1)$) as less than 10\% and greater than 40\%, respectively. 
To achieve these values, we adjust the parameters in the data-generating mechanism in (\ref{eq:dgm}) under DX. Specific parameter values are in the supplement (Section \ref{sec:parvals}), where importantly we have that $\delta_{a,v} \ne 0$. We used 30 clusters with cluster sizes of 100 to 300 with ICCs of 0.01 for both the outcome and validation models. The true ATE is about 17.6\% across all scenarios.

This figure reveals several key insights. Across all four scenarios, our proposed SSW estimator has low empirical bias, but the remaining estimators, IPSW, SSO, and SSW assuming a homogeneous classification model show clear bias. Classification error (which is non-linear) is driving the bias in the SSO estimator. This observation is reinforced by the fact that the magnitude of bias of the SSO estimator increases when the classification error rate increases. For the IPSW estimator, we fit the selection model to match the terms in the data generating mechanism excluding the outcome. Clearly, we cannot regress $V_{ij}$ on $Y_{ij}$ (since $V_{ij}=1$ for observable $Y_{ij}$), so not only is this ATE not estimable in this manner, but also the parameter estimates within the selection model themselves will not necessarily target the parameters of the data generating process unbiasedly due to a lack of collapsibility of the logit link, i.e., the selection model is incorrectly specified when $\delta_{a,v} \ne 0$.  However, unlike the SSO estimator, the direction and magnitude of bias in the IPSW estimator remains relatively similar across all four scenarios (and constant across amount of measurement error) because it does not rely on the error-prone silver-standard outcomes at all. Lastly, mis-specification of the classification model, by falsely assuming conditional independence from treatment and covariates, results in bias for the SSW estimators across all the scenarios. This claim is supported by the observation that bias of this estimator becomes extremely pronounced under the higher rates of misclassification. Under the homogeneous classification assumption, the estimator is also necessarily further away from 0 than the SSO estimator (since the common denominator across treatment factors out), which is not the case for the proposed SSW estimator. 

As expected, the proposed SSW estimator tends to be its most efficient, as measured by its spread, when there is small measurement error with a large validation subset and least efficient when these sizes are switched. Importantly, it has less variability in small samples than the IPSW estimator for both validation subset sizes with small measurement error, underscoring its statistical value even in settings when the IPSW estimator may be applicable. When there is large measurement error, the IPSW estimator has considerably less empirical variance than the SSW estimator, where the latter has to rely on heavy lifting of the classification model for reweighting to correct for misclassification bias. Unfortunately, since the IPSW estimator may be biased under gold-standard outcome-dependent selection, we cannot take advantage of its efficiency and target a combined estimator, namely a convex combination of IPSW and SSW estimators, as used in the previous literature \citep{shen2025integrating,shu2019causal}. While efficiency of the SSW and IPSW estimators are naturally affected by the size of the validation subset, the biased SSO estimator has similar spread across measurement error scenarios because it does not use the gold-standard values at all and thus has roughly the same number of observations. In summary, the proposed SSW estimator displays low bias in small samples for this data generating mechanism, but its empirical variability grows when the amount of measurement error is increased and the proportion of individuals in the validation subset decreases, and its efficiency gains relative to the IPSW estimator will depend on the magnitude of those values.

Finally, although rare, invalid ATE estimates falling outside of $[-1,1]$ (including those producing NA values) were excluded from the simulation results. We report no such failures in Table \ref{tab:sim_table1}, Table \ref{tab:sim_tableapp2}, and Table \ref{tab:sim_tableapp3}. For the resampling methods included in Table \ref{tab:sim_tableapp1}, the average failure rate per simulated data set was nearly zero -- below $0.006\%$ for the cluster bootstrap and below $0.001\%$ for the individual bootstrap across all included scenarios. For Figure~\ref{fig:compareestim}, our SSW estimator experienced failures only under the scenario characterized by large misclassification error and a small selection subset (bottom left panel), and even then, just in $0.4\%$ of iterations. Such pathological estimates (albeit infrequent) arise because we have the most limited amount of data for estimating the most extreme misclassification as is necessary to overcome bias. Notably, using the homogeneous classification model resulted in a failure rate of $1.04\%$ in the bottom left panel and $0.22\%$ in the top left panel, further demonstrating the precariousness of the homogeneity assumption. 

\section{Analysis of ASPIRE}
\label{sec:application}

For the ASPIRE study, we analyze the effect of the nudge+ as compared to the nudge strategy on parent receipt of the complete SAFE Firearm program from clinicians during well-child visits. Recall that the nudge strategy, $A_i=0$, constituted an EHR documentation template incorporated into the standard well-child visit workflow for clinicians, which leads clinicians through prompts to record delivery of both components of the program (i.e., yes or no to ``secure firearm discussed" and ``cable lock offered"). Nudge+, $A_i=1$, supplemented this EHR documentation template with clinic-wide support (e.g., training, planning, ongoing consultation) for clinicians administering the program. While silver-standard clinician-reported measures, $Y^*_{ij}$, were able to be recorded during all visits, these do not directly measure parent receipt and therefore may not correspond to gold-standard parent-reported measures, $Y_{ij}$, which were observed only for survey respondents. 

\begin{table}[t]
\centering

\captionsetup{labelfont=bf,justification=raggedright,singlelinecheck=false}

\caption{Two by two classification tables comparing silver-standard  clinician reports $Y^*_{ij}$ to gold-standard parent reports $Y_{ij}$, stratified by treatment arm (nudge+ and nudge) for the ASPIRE data. Denominators used in parenthetical percentages are the totals within each sub-table, that is the totals under nudge+, nudge, and overall respectively.} 
\begin{tabular}{ccccccccc}
\toprule
 & \multicolumn{2}{c}{\textbf{Nudge+}} &\multicolumn{2}{c}{\textbf{Nudge}}  & \multicolumn{2}{c}{\textbf{Overall}} \\
\midrule
 & $Y^*_{ij}$ = 0 & $Y^*_{ij}$ = 1 & $Y^*_{ij}$ = 0 & $Y^*_{ij}$ = 1 & $Y^*_{ij}$ = 0 & $Y^*_{ij}$ = 1 \\
\midrule
$Y_{ij}$ = 0 & 1062 (48.8\%) & 306 (14.0\%) & 1388 (56.8\%) & 375 (15.3\%) & 2450 (53.0\%) & 681 (14.7\%) \\
$Y_{ij}$ = 1 & 131 (6.0\%) & 679 (31.2\%)  & 287 (11.7\%)  & 393 (16.1\%) & 418 (9.0\%) & 1072 (23.2\%)  \\
\bottomrule
\end{tabular}
\label{tab:aspire_class}
\end{table}

Of the 30 pediatric primary care clinics randomized, 15 sites were assigned the nudge+ treatment and 15 sites were assigned the nudge treatment with 368 clinicians. The number of patient visits totaled 22,317 for nudge+ and 24,989 for nudge. There were 4,621  survey respondents (2,178 for nudge+ and 2,443 for nudge) corresponding to those visits, who reported on whether they believed that their child's clinician delivered both components of SAFE Firearm. In Table \ref{tab:aspire_class}, we display the classification rates in two-way tables, stratified by treatment arm, at the patient visit level. We observe that both marginally and under each treatment individually, misclassification rates are greater when the clinician reported that they had delivered the program, i.e., $Y^*_{ij}=1$ and $Y_{ij}=0$. Moreover, we see heterogeneity across treatment arms, where notably the misclassification rates are higher for the nudge arm, particularly when $Y^*_{ij}=0$ and $Y_{ij}=1$. 

Using our proposed method, we specify a logistic GEE classification model with an independence working correlation, fit on the selection subset. Clinician reporting (silver-standard) is regressed on parent reporting (gold-standard), treatment, and other covariates as defined in the original study \citep{beidas2024implementation} including ethnicity (binary indicator for Hispanic/Latino), gender, clinic volume at enrollment (defined as number of patient-visits in year prior to study), and which of two healthcare systems each clinic (and thus its patients) belongs. The model also includes interactions of treatment with parent reporting, ethnicity, gender, and clinic volume. Choice of the explanatory variables reflects our understanding of the mechanism of classification in these data. For example, we incorporate clinic volume in this way because it might impact the rate of classification errors for such reasons as busier patient schedules or resource accessibility. For illustration in Figure \ref{fig:app1}, we demonstrate the heterogeneity in rates of misclassification by clinic volume, stratified by treatment arm in the ASPIRE data; additionally, this figure shows the number of collected gold-standard outcomes by cluster volume, which appears roughly proportional.  In Table \ref{tab:aspire_an}, we present our proposed estimator, $\hat \tau_{\text{SSW}}$  (\ref{eq:sswestimator}), and we use a cluster-robust variance estimate as in (\ref{eq:crob}) to construct a small-sample corrected $t$-interval (with 23 degrees of freedom). We estimate the ATE for the parent receipt outcome to be $0.111 (-0.013,0.235)$ for nudge+ versus nudge. Our result suggests that there is an estimated increase of approximately $11$ percentage points in the rate of parent receipt of SAFE Firearm under the nudge+ arm as compared to the nudge arm, but this estimate is not statistically significant at $\alpha=0.05$. 

\begin{table}[ht]
\centering
\captionsetup{labelfont=bf,justification=raggedright,singlelinecheck=false}
\caption{Estimation of the average treatment effect (ATE) comparing nudge+ to nudge on parent receipt of the SAFE Firearm program at standard well-child visits in the ASPIRE study. The proposed silver-standard weighting (SSW) estimator is compared to the silver-standard only (SSO) and inverse probability of selection weighting (IPSW) estimators. In ASPIRE, the silver-standard measures are clinician reports, recorded in the EHR at each well-child visit. The gold-standard measures are parent reports, obtained via survey after each visit. Point estimates for the ATE as well as 95\% confidence intervals are reported. The confidence interval based on the SSW estimator is computed using the robust variance expression as demonstrated in Section \ref{sec:estimation}, and non-parametric cluster bootstrap percentile intervals are used with the remaining estimators. One person was removed from the selection subset (0.02\%), and six people in total (0.01\%) had missing values for one of the observed covariates, treatment arm, or silver-standard outcome.}

\begin{tabular}{l c}
\toprule
\textbf{ATE Estimator} & \textbf{Point Estimate (95\% Interval)} \\
\midrule
Silver-standard weighting ($\hat \tau_{\text{SSW}}$)  & 0.111 (-0.013, 0.235) \\
Silver-standard only ($\hat \tau_{\text{SSO}}$)  & 0.131 \hspace{.01cm} (0.017, 0.303) \\
Inverse probability of selection weighting ($\hat \tau_{\text{IPSW}}$) & 0.095 (-0.016, 0.216) \\
\bottomrule
\end{tabular}
\label{tab:aspire_an}
\end{table}

In Table \ref{tab:aspire_an}, we also compare $\hat \tau_{\text{SSW}}$ to the silver-standard only estimator $\hat \tau_{\text{SSO}}$ (\ref{eq:sso_estim}) and the gold-standard subset based IPSW estimator $\hat \tau_{\text{IPSW}}$ (\ref{eq:ipsw_estim}), where the selection model contains the regressors of the classification model less the terms involving parent reporting. For $\hat \tau_{\text{SSO}}$ and $\hat \tau_{\text{IPSW}}$, confidence intervals are obtained using percentiles of the non-parametric cluster bootstrap distributions. While all three point estimates are somewhat similar and are mutually covered by each of the 95\% confidence intervals, we have shown theoretically and in simulations that the SSW estimator protects against bias and serves as the only assured valid estimate for targeting the ATE for the outcome of interest. Interestingly, we see that both the SSW and IPSW estimates are attenuated relative to the SSO estimate, reflecting overly optimistic clinician reporting on delivery in the nudge+ arm. Furthermore, all intervals have relatively comparable widths, but importantly, at three decimal places, the SSW and IPSW intervals cover 0, leading to the same conclusion, whilst the SSO estimator using only clinician reports does not. 

\section{Discussion}
\label{sec:discussion}

Motivated by the challenges associated with ASPIRE, we developed a general framework for identification and estimation of the ATE when there are misclassified binary outcomes and a non-random internal validation subset for cluster-randomized data. The fundamental assumptions enabling non-parametric identification are that potential silver-standard outcomes are causally linked with potential gold-standard outcomes, but only gold-standard outcomes directly impact selection, and that selection is ignorable only when considering the relationship of these potential outcomes given a set of covariates. Estimation of the ATE thus relies on estimating classification probabilities relating the silver-standard to gold-standard outcomes across treatment arms within the internal validation subset, which may incorporate cluster-level and individual-level covariates. Modeling is required if continuous covariates are used, and we opted to represent the classification model using a logistic GEE with an independence working correlation, allowing for effect heterogeneity by treatment. For variance estimation, we appealed to M-estimation theory, constructing stacked unbiased estimating equations by placing the GEE on top of the estimating equations required for finding the component means of our proposed SSW estimator for the ATE. Our approach has the advantage of both reducing to a strictly non-parametric estimator when covariate effects are null and to the iid setting when there is no clustering. Through simulations, we showed that the SSW estimator has low bias in small samples, if correctly accounting for covariates in the classification model, whilst both the naive SSO and IPSW estimators have noticeable bias. Small differences in these estimates do appear in the ASPIRE data, where notably the $95\%$ confidence interval based on the SSW estimator contains 0 unlike with the SSO estimator. In accordance with the previous literature \citep{shu2019causal,shen2025integrating}, we revealed that the SSW estimator can also improve efficiency relative to the IPSW estimator (that averages over values in the validation subset only) under modest measurement error. Finally, we demonstrated that coverage of the $t$-intervals using asymptotic cluster-robust variance estimates is generally close to nominal but may require more aggressive small-sample corrections with larger cluster sizes under non-zero values of ICC. 

There are limitations regarding our method that present fruitful directions for future research. First, we illustrated how to identify the ATE non-parametrically to address misclassification and selection biases. But, for the SSW estimator, which incorporates covariates in the classification model, we fit a GEE that could introduce bias for ATE if it mis-specifies the relationship between the gold- and silver-standard outcomes. In general, effect estimation in the presence of measurement error is sensitive to the classification model, and we saw this bear out in our simulations for our setting. But, while the GEE model specification affords familiarity, computational efficiency, and simple asymptotics,
it is not intrinsic to our identification, and a more flexible modeling approach for hierarchical data can be applied to mitigate the risk of mis-specification. Second, although we wrote out the SSW estimator in a form that is suggestive of inverse probability weighting, our methods have yet to be explicitly extended to the observational data setting. Depending on the design and features of the observational study, care must be taken in re-framing our proposed causal assumptions, particularly if and how the data are clustered \citep{chang2022propensity,ye2024nonparametric}. Moreover, we would need to provide models for an individual's silver-standard outcome and/or propensity score to estimate the ATE. Therefore, we are actively exploring multiply-robust estimators of the ATE, where consistency is maintained if no more than one of the outcome, propensity score, or classification models is mis-specified. Third, our estimator is only valid for binary outcomes. We rely on the implicit expression afforded by the LOTE 
\begin{equation}
 \label{eq:implicit}
    E\left\{E(Y^*_{ij}(a)|Y_{ij}(a),X_{ij})|X_{ij}\right\}=E(Y^*_{ij}(a)|X_{ij}),
\end{equation}
which enables writing the key identification identity in Equation (\ref{eq:eysRHS}). When $Y_{ij}(a)$ is binary, we can represent the distribution of $Y_{ij}(a)|X_{ij}$ in terms of expectations, namely $P(Y_{ij}(a)=1|X_{ij})=E(Y_{ij}(a)|X_{ij})$, without resorting to further assumptions. For continuous outcomes, one simple approach is to assume $E(Y^*_{ij}(a)|Y_{ij}(a),X_{ij})$ is equal (via an identity function) to a function of the regressor variables and is linear in its $Y_{ij}(a)$ arguments; the remaining assumptions of this paper would then enable estimation of this models' parameters. While this is a straightforward extension, identification would no longer be strictly non-parametric. 

There are some features of the ASPIRE data specifically that may also limit the interpretation and application of our method. In this work, we defined our outcome of interest as \textit{parent receipt} of SAFE Firearm as opposed to simply SAFE Firearm delivery. We made this choice because parents may not have correctly recalled whether the program was actually delivered to them by clinicians during standard well-child visits. Therefore, even in using parent-reported outcome measures, we may risk misclassification of true program delivery, leading to recall bias for estimation of the corresponding ATE \citep{althubaiti2016information}. Another important consideration is that clinicians may have varied in how they reported delivery in the EHR and their actions may have also influenced parent perception of program delivery. Although we demonstrated robustness of our method to clinician-induced variation by generating data with clinician random effects in supplementary simulations, we may not only need to explicitly incorporate the hierarchical structure in modeling and variance estimation (even though randomization is at the cluster level) but also collect and adjust for clinician covariates in a new analysis. Similarly, we did not address the possibility that well-child visits within the same family were more likely to have the same reported outcomes (by either clinicians or parents) than visits across different families. Even though failing to account for the full nested structures of our data should not lead to asymptotically biased estimation if our classification mean model remains correctly specified \citep{liang1986longitudinal}, we may see small-sample bias and underestimation in our asymptotic-based variance estimates. Furthermore, since the treatment arms were first assigned to clinics after which clinicians may or may not have administered SAFE Firearm to parents, the effect of nudge+ versus nudge on the parent receipt outcome may have been mediated by clinician quality. While the ASPIRE data lack sufficient information about clinicians to enable a mediation analysis, exploring clinician mediating effects in a future study might offer additional insights beyond those captured by total effects alone \citep{vanderweele2015explanation}. 

In conclusion, our proposed causal framework and subsequent SSW estimator for the ATE have addressed key challenges related to measurement error and selection mechanisms in cluster-randomized trials.  When the selection mechanism is complicated or not well-understood, the SSW estimator may outperform estimators that rely on specifying a selection model.  In addition, our use of the SSW estimator revealed new insights about the ASPIRE trial as compared to only using silver-standard outcomes. Our approach demonstrates how to take advantage of the strengths of different measures for the same outcome -- as are often available in studies that rely on EHR data -- to ultimately target an effect measure of interest.

\section*{Funding}

This work was supported by funding from the National Institute of Mental Health (R01 MH123491 and R01 MH123491-06S1, to Rinad Beidas, PhD and Kristin Linn, PhD respectively).

\section*{Acknowledgments}

We thank the Site Principal Investigators at the two health systems: Brian Ahmedani, PhD (Henry Ford Health System) and Jennifer Boggs, PhD (Kaiser Permanente Colorado Health System). We acknowledge Christina Johnson, MPH (Northwestern University) for overseeing operations of the parent grant and supplement that funded our work. We recognize Katy Bedjeti, MSW, MS (Northwestern University) for assembling the ASPIRE data.  We thank Owen Yoo (undergraduate studies at University of Michigan) for independently replicating our simulation studies.  

\bibliographystyle{plainnat}
\bibliography{Bibliography}

\pagebreak

\appendix
\setcounter{table}{0}
\input{ASPIRE_manu_clean_draftsupp}

\end{document}

%% file: ASPIRE_manu_clean_draftsupp.tex
\renewcommand{\thefigure}{S\arabic{figure}} 
\setcounter{figure}{0} 

\renewcommand{\thetable}{S\arabic{table}} 
\setcounter{table}{0} 

\section{Identification}
\label{sec:identification}

Recall the following identification assumptions:

\begin{itemize}

\item[\textbf{IA1}] Cluster-level SUTVA: Under $A_i=a \in \{0,1\}$, a) $Y_{ij}^*=Y_{ij}^*(a)$; b) $V_{ij}=V_{ij}(a)$; and c) $Y_{ij}=Y_{ij}(a)$ when $V_{ij}(a)=1$.

\item[\textbf{IA2}] Randomization: For $a=0,1$, $\{Y^*_i(a),Y_i(a),V_i(a),L_i,U_i\} \perp A_i$ where $0<P(A_i=1)<1$.

\item[\textbf{IA3}] Classification ignorability: For $a=0,1$ and all individuals, $Y_{ij}^*(a) \perp V_{ij}(a)|Y_{ij}(a),X_{ij}$. 

\begin{itemize}
    \item[\textbf{IA3*}] Stronger classification ignorability: For $a=0,1$ and all individuals, $\{Y_{ij}^*(a),Y_{ij}(a)\} \perp V_{ij}(a)|X_{ij}$. 
\end{itemize}

\item [\textbf{IA4}] Differential classification: For $a=0,1$ and all individuals, 
\begin{equation*}
P(Y^*_{ij}(a)=1|V_{ij}(a)=1,Y_{ij}(a)=1,X_{ij}) \ne P(Y^*_{ij}(a)=1|V_{ij}(a)=1,Y_{ij}(a)=0,X_{ij})
\end{equation*}
\end{itemize}

From IA1, IA2, and selection ignorability $Y_{ij}(a) \perp V_{ij}(a)|X_{ij}$ via IA3*, we demonstrate that the ATE is identified with the inverse probability of selection weighting (IPSW) expression, $\tau_{\text{IPSW}}=\mu_{\text{IPSW}}(1)-\mu_{\text{IPSW}}(0)$, in Equation (\ref{eq:ipsw}). We proceed as follows 
\begin{align*}
E\{Y_{ij}(a)\}&=E\left\{E(Y_{ij}(a)|X_{ij})\right\}\\
&=E\left\{E(Y_{ij}(a)|V_{ij}(a)=1,X_{ij})\right\} \hspace{.3cm} \text{(by selection ignorability per IA3*)}\\
&=E\left\{E(Y_{ij}(a)|V_{ij}(a)=1,A_{ij}=a,X_{ij})\right\} \hspace{.3cm} \text{(by IA2)}\\
&=E\left\{E(Y_{ij}|V_{ij}=1,A_{ij}=a,X_{ij})\right\} \hspace{.3cm} \text{(by IA1)}\\
&=E\left\{E\left(\frac{I(A_{ij}=a)V_{ij}Y_{ij}}{E(V_{ij}|A_{ij}=a,X_{ij})P(A_{ij}=a)}\biggr|X_{ij}\right)\right\}\\
&=E\left\{\frac{I(A_{ij}=a)V_{ij}Y_{ij}}{E(V_{ij}|A_{ij}=a,X_{ij})P(A_{ij}=a)}\right\}=\mu_{\text{IPSW}}(a) \hspace{.3cm} \text{(by LOTE)}
\end{align*}
with $V_{ij}Y_{ij}=0$ if $V_{ij}=0$. Since $a \in \{0,1\}$ is arbitrary, the ATE is identified with $\tau_{\text{IPSW}}$ using the gold-standard subset only, according to assumptions IA1, IA2, and selection ignorability. \\

\noindent Under IA1-IA4, we demonstrate Proposition \ref{identity1} in Equation (\ref{eq:ssw}), which says the ATE is identified with the expression $\tau_{\text{SSW}}=\mu_{\text{SSW}}(1)-\mu_{\text{SSW}}(0)$. The proof proceeds as follows:
\begin{align*}
E(Y_{ij}^*(a)|X_{ij})\stackrel{\text{IA2}}{=}E(Y_{ij}^*(a)|A_{ij}=a,X_{ij})&\stackrel{\text{IA1}}{=}E(Y_{ij}^*I(A_{ij}=a)|X_{ij})\left[P(A_{ij}=a|X_{ij})\right]^{-1}\\
&\stackrel{\text{IA2}}{=}E(Y_{ij}^*I(A_{ij}=a)|X_{ij})\left[P(A_{ij}=a)\right]^{-1}.
\end{align*}
Simultaneously by law of total expectation (LOTE),
\begin{align*}
E(Y_{ij}^*(a)|X_{ij})&=E\left\{Y_{ij}^*(a)|Y_{ij}(a)=1,X_{ij}\right\}E(Y_{ij}(a)|X_{ij})+E\left\{Y_{ij}^*(a)|Y_{ij}(a)=0,X_{ij}\right\}\left[1-E(Y_{ij}(a)|X_{ij})\right]\\
&=E\left\{Y_{ij}^*(a)|V_{ij}(a)=1,Y_{ij}(a)=1,X_{ij}\right\}E(Y_{ij}(a)|X_{ij})\\
&\quad +E\left\{Y_{ij}^*(a)|V_{ij}(a)=1,Y_{ij}(a)=0,X_{ij}\right\}\left[1-E(Y_{ij}(a)|X_{ij})\right] \hspace{.3cm} \text{(by IA3, IA4)} \\
&=E\left\{Y_{ij}^*(a)|V_{ij}(a)=1,Y_{ij}(a)=1,A_{ij}=a,X_{ij}\right\}E(Y_{ij}(a)|X_{ij})\\
&\quad +E\left\{Y_{ij}^*(a)|V_{ij}(a)=1,Y_{ij}(a)=0,A_{ij}=a,X_{ij}\right\}\left[1-E(Y_{ij}(a)|X_{ij})\right] \hspace{.3cm} \text{(by IA2)}\\
&=E\left\{Y_{ij}^*|V_{ij}=1,Y_{ij}=1,A_{ij}=a,X_{ij}\right\}E(Y_{ij}(a)|X_{ij})\\
&\quad +E\left\{Y_{ij}^*|V_{ij}=1,Y_{ij}=0,A_{ij}=a,X_{ij}\right\}\left[1-E(Y_{ij}(a)|X_{ij})\right] .\hspace{.3cm} \text{(by IA1)}
\end{align*}
Setting these two expressions for $E(Y_{ij}^*(a)|X_{ij})$ equal, we have 
\begin{align*}
E\left(Y_{ij}^*I(A_{ij}=a)|X_{ij}\right)\left[P(A_{ij}=a)\right]^{-1} &=E\left\{Y_{ij}^*|V_{ij}=1,Y_{ij}=1,A_{ij}=a,X_{ij}\right\}E(Y_{ij}(a)|X_{ij})\\
&\quad +E\left\{Y_{ij}^*|V_{ij}=1,Y_{ij}=0,A_{ij}=a,X_{ij}\right\}\left[1-E(Y_{ij}(a)|X_{ij})\right]. \hspace{.3cm}
\end{align*}
Re-arranging and dividing through (where IA4 has prevented division by 0), we have
\begin{align*}
E(Y_{ij}(a)|X_{ij})&=\frac{E\left(Y_{ij}^*I(A_{ij}=a)|X_{ij}\right)\left[P(A_{ij}=a)\right]^{-1}-E\left\{Y_{ij}^*|V_{ij}=1,Y_{ij}=0,A_{ij}=a,X_{ij}\right\}}{E\left\{Y_{ij}^*|V_{ij}=1,Y_{ij}=1,A_{ij}=a,X_{ij}\right\}-E\left\{Y_{ij}^*|V_{ij}=1,Y_{ij}=0,A_{ij}=a,X_{ij}\right\}}\\
&=E\left(\frac{Y_{ij}^*I(A_{ij}=a)-E\left\{Y_{ij}^*|V_{ij}=1,Y_{ij}=0,A_{ij}=a,X_{ij}\right\}P(A_{ij}=a)}{P(A_{ij}=a)\left[E\left\{Y_{ij}^*|V_{ij}=1,Y_{ij}=1,A_{ij}=a,X_{ij}\right\}-E\left\{Y_{ij}^*|V_{ij}=1,Y_{ij}=0,A_{ij}=a,X_{ij}\right\}\right]}\bigg |X_{ij}\right).
\end{align*}
Thus, taking the expectation over the distribution of $X_{ij}$ of the right and left hand sides, we get 
\begin{align*}E\{Y_{ij}(a)\}&=E\left\{\frac{Y_{ij}^*I(A_{ij}=a)-E\left\{Y_{ij}^*|V_{ij}=1,Y_{ij}=0,A_{ij}=a,X_{ij}\right\}P(A_{ij}=a)}{P(A_{ij}=a)\left[E\left\{Y_{ij}^*|V_{ij}=1,Y_{ij}=1,A_{ij}=a,X_{ij}\right\}-E\left\{Y_{ij}^*|V_{ij}=1,Y_{ij}=0,A_{ij}=a,X_{ij}\right\}\right]}\right\}\\
&=E\left\{\frac{Y_{ij}^*I(A_{ij}=a)-P(A_{ij}=a)p^v_{ij}(0,a,X_{ij})}{P(A_{ij}=a)[p^v_{ij}(1,a,X_{ij})-p^v_{ij}(0,a,X_{ij})]}\right\}=\mu_{\text{SSW}}(a).
\end{align*}
Since $a \in \{0,1\}$ is arbitrary, the ATE is identified with $\tau_{\text{SSW}}$ as proposed according to IA1-IA4. 

\section{Unbiased Estimating Equations and Variance Estimation}
\label{sec:uee}

Stack the following components to define $m_i(\lambda)$ for cluster $i$, which enable constructing UEEs (indexed by cluster) whose solution is $\hat \lambda$. We can then  immediately find the inner/meat matrix $\frac{1}{m}\sum_{i=1}^{m}m_{i}(\widehat \lambda)m_{i}(\widehat \lambda)^T$ of the sandwich variance estimator $\hat V_{\hat \lambda}$ defined in Equation (\ref{eq:crob}) of the main text.

\begin{enumerate}
    \item Solving for $\hat \theta$ (through the GEE) 

    $$m_{i,1}(\lambda)=\sum_{j:V_{ij}=1}D_{ij}\left(Y_{ij}^*-\frac{\exp(D_{ij}^T\theta)}{1+\exp(D_{ij}^T\theta)}\right)$$

    where recall that $D^T_{ij}=(1,Y_{ij},A_{ij},Y_{ij}A_{ij},W_{ij}^T)$ is a row of the design  matrix $D_i$ for cluster $i$. $W_{ij}$ includes any necessary remaining regressors from the components of $X_{ij}$ and their interactions.

    \item Solving for $\hat \pi$

    $$m_{i,2}(\lambda)=\sum_{j=1}^{n_i}(A_{ij}-\pi)$$ 

    \item Solving for $\hat \mu_{\text{SSW}}(1)$

    $$m_{i,3}(\lambda)=\sum_{j=1}^{n_i} \left(\frac{A_{ij}Y^*_{ij}- \pi p^v_{ij}(D_{ij,01};\theta)}{ \pi [ p^v_{ij}(D_{ij,11};\theta)- p^v_{ij}(D_{ij,01};\theta)]}-\mu_{\text{SSW}}(1)\right)$$

    \item Solving for $\hat \mu_{\text{SSW}}(0)$

    $$m_{i,4}(\lambda)=\sum_{j=1}^{n_i}\left(\frac{(1-A_{ij})Y^*_{ij}-(1-\pi) p^v_{ij}(D_{ij,00};\theta)}{(1- \pi) [p^v_{ij}(D_{ij,10};\theta)- p^v_{ij}(D_{ij,00};\theta)]}-\mu_{\text{SSW}}(0)\right)$$

    \end{enumerate}

\noindent $D_{ij,ya}$ is a row of the design matrix for person $ij$ where $Y_{ij}$ is set to $y \in \{0,1\}$ and $A_{ij}$ is set to $a \in \{0,1\}$. The components of the derivative (Jacobian) $\frac{\partial m_{i}}{\partial \lambda^T}$ are needed to find the bread/outer matrix $\left[\frac{1}{m}\sum_{i=1}^{m} \left \{\frac{\partial m_{i}(\widehat \lambda)}{\partial \lambda^T}\right \} \right]^{-1}$, where we plug in $\hat \lambda$ and use \texttt{R} to compute the inverse. They are as follows:
    \begin{enumerate}

        \item $\sum_{i=1}^{m}\frac{\partial m_{i,1}(\lambda)}{\partial \lambda^T}$

        \begin{enumerate}
            \item       $\sum_{i=1}^{m}\frac{\partial m_{i,1}(\lambda)}{\partial \theta^T}=$
            $$-\sum_{j:V_{ij}=1}\frac{\exp(D_{ij}^T\theta)}{[1+\exp(D_{ij}^T\theta)]^2}D_{ij}D_{ij}^T$$
            \item The rest of the partial derivatives for this submatrix are 0. 
            
        \end{enumerate}

        \item $\sum_{i=1}^{m}\frac{\partial m_{i,2}(\lambda)}{\partial \lambda^T}$

        Simply $\sum_{i=1}^{m}\frac{\partial m_{i,2}(\lambda)}{\partial \pi}=-N$. The rest of the partial derivatives for this row submatrix are 0.

        \item $\sum_{i=1}^{m}\frac{\partial m_{i,3}(\lambda)}{\partial \lambda^T}$

        \begin{enumerate}

            \item $\left(\sum_{i=1}^{m}\frac{\partial m_{i,3}(\lambda)}{\partial \theta^T}\right)^T=$
            \begin{align*}&\sum_{j=1}^{n_i}\frac{-\pi^2 D_{ij,01}p^v_{ij}(D_{ij,01};\theta)(1-p^v_{ij}(D_{ij,01};\theta))(p^v_{ij}(D_{ij,11};\theta)-p^v_{ij}(D_{ij,01};\theta))}{\pi^2[p^v_{ij}(D_{ij,11};\theta)-p^v_{ij}(D_{ij,01};\theta)]^2}\\
            &-\bigg [\frac{\pi (A_{ij}Y^*_{ij}-\pi p^v_{ij}(D_{ij,01};\theta))}{\pi^2[p^v_{ij}(D_{ij,11};\theta)-p^v_{ij}(D_{ij,01};\theta)]^2} \times
            \\& \left\{D_{ij,11}p^v_{ij}(D_{ij,11};\theta)(1-p^v_{ij}(D_{ij,11};\theta))-D_{ij,01}p^v_{ij}(D_{ij,01};\theta)(1-p^v_{ij}(D_{ij,01};\theta))\right\}\bigg]
            \end{align*}

            \end{enumerate}

           We allow for left and right scalar multiplication to a vector. We leave this expression (as well as 4(a)) unsimplified for interpretability. 

            \begin{enumerate}

            \item[(b)] $\sum_{i=1}^{m}\frac{\partial m_{i,3}(\lambda)}{\partial \pi}=$

            $$\sum_{j=1}^{n_i}\frac{-A_{ij}Y^*_{ij}}{\pi^2[p^v_{ij}(D_{ij,11};\theta)-p^v_{ij}(D_{ij,01};\theta)]}$$

            \item[(c)] $\sum_{i=1}^{m}\frac{\partial m_{i,3}(\lambda)}{\partial \mu_{\text{SSW}}(1)}=-N$ and $\sum_{i=1}^{m}\frac{\partial m_{i,3}(\lambda)}{\partial \mu_{\text{SSW}}(0)}=0$
            
        \end{enumerate}

            \item $\sum_{i=1}^{m}\frac{\partial m_{i,4}(\lambda)}{\partial \lambda^T}$

        \begin{enumerate}

            \item $\left(\sum_{i=1}^{m}\frac{\partial m_{i,4}(\lambda)}{\partial \theta^T}\right)^T=$
            \begin{align*}&\sum_{j=1}^{n_i}\frac{-(1-\pi)^2 D_{ij,00}p^v_{ij}(D_{ij,00};\theta)(1-p^v_{ij}(D_{ij,00};\theta))(p^v_{ij}(D_{ij,10};\theta)-p^v_{ij}(D_{ij,00};\theta))}{(1-\pi)^2[p^v_{ij}(D_{ij,10};\theta)-p^v_{ij}(D_{ij,00};\theta)]^2}\\
            &-\bigg[\frac{(1-\pi)((1-A_{ij})Y^*_{ij}-(1-\pi) p^v_{ij}(D_{ij,00};\theta))}{(1-\pi)^2[p^v_{ij}(D_{ij,10};\theta)-p^v_{ij}(D_{ij,00};\theta))]^2} \times \\
            &\left\{D_{ij,10}p^v_{ij}(D_{ij,10};\theta)(1-p ^v_{ij}(D_{ij,10};\theta))-D_{ij,00}p^v_{ij}(D_{ij,00};\theta)(1-p^v_{ij}(D_{ij,00};\theta))\right\}\bigg]
            \end{align*}
            
            \item $\sum_{i=1}^{m}\frac{\partial m_{i,4}(\lambda)}{\partial \pi}=$

            $$\sum_{j=1}^{n_i}\frac{(1-A_{ij})Y^*_{ij}}{(1-\pi)^2[p^v_{ij}(D_{ij,10};\theta)-p^v_{ij}(D_{ij,00};\theta)]}$$
            
            \item $\sum_{i=1}^{m}\frac{\partial m_{i,4}(\lambda)}{\partial \mu_{\text{SSW}}(1)}=0$ and $\sum_{i=1}^{m}\frac{\partial m_{i,4}(\lambda)}{\partial \mu_{\text{SSW}}(0)}=-N$

        \end{enumerate}

    \end{enumerate}

\section{Parameter Values for Figure 2}
\label{sec:parvals}

The format of the data-generating mechanism is as in setup (\ref{eq:dgm}) of the main text
\begin{align*}
    & Y_{ij}(a) \sim \Bern\left\{\expit(\psi_{ij}(a,X_{ij},b_{i,y}) \right\} & &\psi_{ij}(a,X_{ij},b_{i,y})=\alpha_{a,y}+\beta_{a,y}^TX_{ij}+b_{i,y} \nonumber\\
    & Y^*_{ij}(a) \sim \Bern\left\{\expit(\rho_{ij}(a,Y_{ij}(a),X_{ij}) \right\} & &\rho_{ij}(a,Y_{ij}(a),X_{ij}) = \alpha_{a,y^*}+\beta_{a,y^*}^{T}X_{ij}+\delta_{a,y^*}Y_{ij}(a)\\
    & V_{ij}(a) \sim \Bern\left\{\expit(\varphi_{ij}(a,Y_{ij}(a),X_{ij},b_{i,v}) \right\} & & \varphi_{ij}(a,Y_{ij}(a),X_{ij},b_{i,v}) = \alpha_{a,v}+\beta_{a,v}^{T}X_{ij}+\delta_{a,v}Y_{ij}(a)+b_{i,v} \nonumber
\end{align*}
the specific parameter values for Figure \ref{fig:compareestim} are
\begin{align*}
(\alpha_{a,y},\beta^T_{a,y})&=(-1+0.75a,0.15,0.2,0.15,-0.15) \\ \nonumber 
(\alpha_{a,y^*},\beta^{T}_{a,y^*},\delta_{a,y^*})&=  \begin{cases}
    (-2-0.75a,-0.55+0.2a,-0.35+0.1a,0.15,-0.1,4+1.75a)  & \text{SME}\\
    (-0.25+0.05a,-0.5+0.15a,-0.35+0.1a,0.15,0,0.7+0.25a)   & \text{LME}
  \end{cases}\\ \nonumber
(\alpha_{a,v},\beta^T_{a,v},\delta_{a,v})&=
\begin{cases}
    (-0.25+0.1a,-0.75,-0.75,-0.75,0.15,0.15-0.3a) & \text{SV}\\
    (0.7-0.25a,-0.5,-0.5,-0.5,0.1,0.15-0.3a)   & \text{LV}
  \end{cases}
\end{align*}
where S indicates small, L indicates large, ME indicates measurement error (misclassification), and V indicates size of the validation set.

\section{Additional Tables of Simulations}
\label{sec:addtables}

We note that code for producing all tables and figures in the main text and supplement (relating to simulations and the data analysis) can be found at \url{https://github.com/abcdane1/mesbestim} in the folder \texttt{TabsandFigures\_SimsandData}.

\begin{itemize}
    \item[S1] Table \ref{tab:sim_tableapp1} reports the coverage
    of competing variance estimation techniques for the SSW estimator  under the data generation for Table \ref{tab:sim_table1}. These competing techniques include: non-parametric cluster bootstrap \citep{field2007bootstrapping}, individual sandwich variance estimator, and non-parametric individual bootstrap. There are 1000 bootstrap iterations for all bootstrap approaches, and intervals are derived from the percentiles of the bootstrap distributions.
        \begin{itemize}
            \item Note, due to computational burden, we only consider the bootstrap methods for cluster sizes of 100 to 300 individuals. This setting was sufficient to illustrate the compatibility (albeit with small differences) of the bootstrap methods to their asymptotic counterparts.  
        \end{itemize}
    \item[S2] Table \ref{tab:sim_tableapp2}
    adds variation due to clinician within site via a clinician random effect in the main data generating mechanism. 
    \item[S3] Table \ref{tab:sim_tableapp3} adds cluster-level correlation via a cluster random effect within the classification model in the main data generating mechanism. 
\end{itemize}

\subsection{Table S1}
\begin{table}[ht]
    \centering
    \captionsetup{labelfont=bf,justification=raggedright,singlelinecheck=false}

    \caption{We compare the empirical coverage of approximate 95\% confidence intervals from different variance estimation methods as a supplement to Table \ref{tab:sim_table1}. Variance estimates are reported alongside the empirical variance for reference. Four methods are compared: cluster-robust sandwich variance with small-sample corrected $t$-interval (C-Robust$^*$), non-parametric cluster-level bootstrap (C-Bootstrap), individual sandwich variance estimate (I-Robust), and non-parametric individual-level bootstrap (I-Bootstrap). 1000 bootstrap iterations are used for all bootstrap methods, and intervals are derived from the percentiles of the bootstrap distributions. Only asymptotic variances are compared for cluster sizes of 500 to 1000 individuals due to computational burden.}
    
    \begin{tabular}{>{\centering\arraybackslash}p{1.3cm} 
                    >{\centering\arraybackslash}p{1.3cm}| 
                    >{\centering\arraybackslash}p{1.2cm} 
                    >{\centering\arraybackslash}p{1.2cm} 
                    >{\centering\arraybackslash}p{1.2cm}
                    >{\centering\arraybackslash}p{1.2cm} 
                    >{\centering\arraybackslash}p{1.2cm} 
                    >{\centering\arraybackslash}p{1.2cm} 
                    >{\centering\arraybackslash}p{1.2cm} 
                    >{\centering\arraybackslash}p{1.2cm}} 
    \toprule
    \multicolumn{2}{c|}{\textbf{Scenarios}} & 
    \multicolumn{2}{c}{\textbf{C-Robust$^*$}} & 
    \multicolumn{2}{c}{\textbf{C-Bootstrap}} & 
    \multicolumn{2}{c}{\textbf{I-Robust}} & 
    \multicolumn{2}{c}{\textbf{I-Bootstrap}} \\
    \cmidrule(lr){1-2} \cmidrule(lr){3-4} \cmidrule(lr){5-6} \cmidrule(lr){7-8} \cmidrule(lr){9-10}
    \thead{Model} & \thead{Empirical \\ Variance} & 
    \thead{Variance} & \thead{Coverage} & 
    \thead{Variance} & \thead{Coverage} & 
    \thead{Variance} & \thead{Coverage} & 
    \thead{Variance} & \thead{Coverage} \\
  \midrule
    \multicolumn{10}{c}{\textit{True Classification Model Excludes Covariates}} \\     
    \addlinespace
    \multicolumn{10}{l}{ICCs: 0.010; Cluster Sizes: $\mathcal{U}\{100,300\}$}\\
    \midrule
    Model 1 & 0.004 & 0.003 & 94.3\% & 0.004 & 93.5\% & 0.003 & 94.6\% & 0.004 & 93.7\% \\
    Model 2 & 0.003 & 0.003 & 94.2\% & 0.003 & 93.3\% & 0.002 & 93.8\% & 0.003 & 93.4\% \\
    \addlinespace
    \multicolumn{10}{l}{ICCs: 0.010; Cluster Sizes: $\mathcal{U}\{500,1000\}$} \\
    \midrule
    Model 1 & 0.001 & 0.001 & 93.6\% & - & - & 0.001 & 90.3\% & - & - \\
    Model 2 & 0.001 & 0.001 & 93.7\% & - & - & 0.001 & 89.1\% & - & - \\
    \addlinespace
    \multicolumn{10}{l}{ICCs: 0.100; Cluster Sizes: $\mathcal{U}\{100,300\}$}\\
    \midrule
    Model 1 & 0.006 & 0.005 & 93.9\% & 0.006 & 93.8\% & 0.003 & 85.2\% & 0.003 & 85.1\% \\
    Model 2 & 0.005 & 0.005 & 94.4\% & 0.005 & 93.6\% & 0.002 & 81.9\% & 0.002 & 81.4\% \\
    \addlinespace
    \multicolumn{10}{l}{ICCs: 0.100; Cluster Sizes: $\mathcal{U}\{500,1000\}$} \\
    \midrule
    Model 1 & 0.003 & 0.003 & 94.2\% & - & - & 0.001 & 65.4\% & - & - \\
    Model 2 & 0.003 & 0.003 & 94.1\% & - & - & 0.001 & 60.6\% & - & - \\
    \bottomrule
    \addlinespace
    \multicolumn{10}{c}{\textit{True Classification Model Includes Covariates}} \\ 
    \addlinespace
    \multicolumn{10}{l}{ICCs: 0.010; Cluster Sizes: $\mathcal{U}\{100,300\}$}\\
    \midrule
    Model 1 & 0.004 & 0.003 & 94.2\% & 0.004 & 93.5\% & 0.003 & 94.3\% & 0.004 & 94.0\% \\
    Model 2 & 0.003 & 0.003 & 79.4\% & 0.003 & 75.2\% & 0.003 & 77.2\% & 0.003 & 74.9\% \\
    \addlinespace
    \multicolumn{10}{l}{ICCs: 0.010; Cluster Sizes: $\mathcal{U}\{500,1000\}$} \\
    \midrule
    Model 1 & 0.001 & 0.001 & 94.0\% & - & - & 0.001 & 91.1\% & - & - \\
    Model 2 & 0.001 & 0.001 & 50.4\% & - & - & 0.001 & 37.2\% & - & - \\
    \addlinespace
    \multicolumn{10}{l}{ICCs: 0.100; Cluster Sizes: $\mathcal{U}\{100,300\}$} \\
    \midrule
    Model 1 & 0.006 & 0.005 & 94.1\% & 0.006 & 93.8\% & 0.003 & 84.9\% & 0.003 & 84.7\% \\
    Model 2 & 0.005 & 0.005 & 87.2\% & 0.005 & 84.9\% & 0.003 & 70.1\% & 0.003 & 68.8\% \\
    \addlinespace
    \multicolumn{10}{l}{ICCs: 0.100; Cluster Sizes: $\mathcal{U}\{500,1000\}$} \\
    \midrule
    Model 1 & 0.003 & 0.003 & 93.9\% & - & - & 0.001 & 65.9\% & - & - \\
    Model 2 & 0.003 & 0.003 & 83.1\% & - & - & 0.001 & 42.1\% & - & - \\
    \bottomrule
\end{tabular}
    \caption*{* indicates that small-sample (of clusters) correction used for confidence intervals. Model 1 - includes covariates in fitted classification model. Model 2 - does not include covariates in fitted classification model. Corrected coverage - coverage of $t$-intervals with $m-7$ degrees of freedom. ICC - intracluster correlation coefficient.}
    \label{tab:sim_tableapp1}
\end{table}

\FloatBarrier

\subsection{Table S2}

The format of the data generating mechanism for Table S2 is as follows:
\begin{align*}
\label{eq:dgmwdoc}
    & Y_{ijk}(a) \sim \Bern\left\{\expit(\psi_{ijk}(a,X_{ijk},b_{i,y},b_{ij,y}) \right\} & &\psi_{ijk}(a,X_{ijk},b_{i,y},b_{ij,y})=\alpha_{a,y}+\beta_{a,y}^TX_{ijk}+b_{i,y}+b_{ij,y}\\
    & Y^*_{ijk}(a) \sim \Bern\left\{\expit(\rho_{ijk}(a,Y_{ijk}(a),X_{ijk}) \right\} & &\rho_{ijk}(a,Y_{ijk}(a),X_{ijk}) = \alpha_{a,y^*}+\beta_{a,y^*}^{T}X_{ijk}+\delta_{a,y^*}Y_{ijk}(a) \nonumber \\
    & V_{ijk}(a) \sim \Bern\left\{\expit(\varphi_{ijk}(a,Y_{ijk}(a),X_{ijk},b_{i,v},b_{ij,v}) \right\} & & \varphi_{ijk}(a,Y_{ijk}(a),X_{ijk},b_{i,v},b_{ij,v}) = \{ \alpha_{a,v}+\beta_{a,v}^{T}X_{ijk}\\& & &+\delta_{a,v}Y_{ijk}(a)+b_{i,v}+b_{ij,v}\} \nonumber
\end{align*}
where $i=1,...,m$ indexes sites, $j=1,...,l_i$ indexes clinicians with $l_i$ clinicians in site $i$, and $k=1,...,n_{ij}$ indexes individual units with $n_{ij}$ individuals for clinician $j$ in site $i$. Therefore, for individual level variables, we are re-indexing by $ijk$; any site or clinician level values will be fixed with respect to index. Thus, we would technically be targeting, $\text{ATE}:=P(Y_{ijk}(1)=1)-P(Y_{ijk}(0)=1)$, but our assumptions and modeling would not be explicitly accounting for this nested structure. We define the fixed effects parameters and random effects $b_{i,y}$, $b_{i,v}$ as in (\ref{eq:dgm}) and (\ref{eq:paramst1}) for Table \ref{tab:sim_table1}. We define $b_{ij,y} \overset{\text{iid}}{\sim} N(0,\sigma^2_{bc,y})$, and $b_{ij,v} \overset{\text{iid}}{\sim} N(0,\sigma^{2}_{bc,v})$, which are independent of fixed effects and cluster-level random effects. Using the latent response representation \citep{eldridge2009intra}, we define within-site within-clinician correlation (WSWC) and within-site across-clinician correlation (WSAC) \citep{arnup2017understanding} as
\begin{align*}
\text{WSWC}=\frac{\sigma^{2}_{b,(\cdot)}+\sigma^{2}_{bc,(\cdot)}}{\sigma^{2}_{b,(\cdot)}+\sigma^{2}_{bc,(\cdot)}+\pi^2/3} & & \text{WSAC}=\frac{\sigma^{2}_{b,(\cdot)}}{\sigma^{2}_{b,(\cdot)}+\sigma^{2}_{bc,(\cdot)}+\pi^2/3}
\end{align*}
where the notation $(\cdot)$ is meant for general representation. We solve a system of equations such that WSWC is 0.01 or 0.1 and WSAC is 0.005 or 0.05 respectively. We vary $l_i$ uniformly between 2 to 3 such that the number of clinicians per site is relatively small as with ASPIRE (but with less variability).

\begin{table}[ht]
    \centering
    \captionsetup{labelfont=bf,justification=raggedright,singlelinecheck=false}
    
    \caption{Evaluation of the small-sample performance of our proposed estimation method under differential misclassification with a non-random internal validation set. We generated data for 30 clusters as described in Section \ref{sec:simstudy} with reference Equations (\ref{eq:dgm}) and (\ref{eq:paramst1}) \textit{adding within-clinician correlation via random effects}. 8 scenarios are reported which have two levels of WSWC=0.01 or 0.1 and WSAC=0.005 or 0.05 (resp.), two levels of cluster sizes, $\mathcal{U}\{100,300\}$ or $\mathcal{U}\{500,100\}$, and exclusion or inclusion of covariates in the classification model generating silver-standard outcomes. Empirical bias and coverage of approximate  95\% normal intervals and small-sample corrected $t$-intervals using the cluster-robust (CR) sandwich variance are reported. The model-based CR variances are compared to the empirical variances for reference. Model 1 which accounts for covariates in fitting a classification model is juxtaposed against Model 2 which has no covariates, i.e., Model 2 estimates classification probabilities non-parametrically.}
    
   \begin{tabular}{>{\centering\arraybackslash}p{1.5cm} 
                >{\centering\arraybackslash}p{1.5cm} 
                >{\centering\arraybackslash}p{1.5cm} 
                >{\centering\arraybackslash}p{1.5cm} 
                >{\centering\arraybackslash}p{1.5cm} 
                >{\centering\arraybackslash}p{1.5cm} 
                >{\centering\arraybackslash}p{1.5cm}}
    \toprule
    \textbf{\thead{Model}} & \textbf{\thead{True \\ ATE}} & \textbf{\thead{Empirical \\ Bias}} & \textbf{\thead{Empirical \\ Variance}} & \textbf{\thead{Model\\ Variance}} & \thead{\textbf{Coverage}} & \textbf{\thead{Corrected \\ Coverage}} \\
    \midrule
    \multicolumn{7}{c}{\textit{True Classification Model Excludes Covariates}} \\     
    \addlinespace
    \multicolumn{7}{l}{WSWCs: 0.010; WSACs: 0.005; Cluster Sizes: $\mathcal{U}\{100,300\}$}\\
    \midrule
    Model 1 & 0.176 & -0.004 & 0.003 & 0.003 & 92.2\% & 93.8\% \\
    Model 2 & 0.176 & -0.001 & 0.003 & 0.002 & 92.5\% & 93.9\% \\
    \addlinespace
    \multicolumn{7}{l}{WSWCs: 0.010; WSACs: 0.005; Cluster Sizes: $\mathcal{U}\{500,1000\}$} \\
    \midrule
    Model 1 & 0.176 & -0.001 & 0.001 & 0.001 & 92.4\% & 93.8\% \\
    Model 2 & 0.176 &  -0.000 & 0.001 & 0.001 & 92.6\% & 93.7\% \\
    \addlinespace
    \multicolumn{7}{l}{WSWCs: 0.100; WSACs: 0.050; Cluster Sizes: $\mathcal{U}\{100,300\}$}\\
    \midrule
    Model 1 & 0.165 & -0.004 & 0.005 & 0.004 & 92.1\% & 93.6\% \\
    Model 2 & 0.165 & -0.002 & 0.004 & 0.004 & 92.3\% & 93.8\% \\
    \addlinespace
    \multicolumn{7}{l}{WSWCs: 0.100; WSACs: 0.050; Cluster Sizes: $\mathcal{U}\{500,1000\}$} \\
    \midrule
    Model 1 & 0.165 & -0.000 & 0.003 & 0.002 & 92.4\% & 93.7\% \\
    Model 2 & 0.165 & 0.000 & 0.002 & 0.002 & 92.3\% & 93.6\% \\
    \bottomrule
    \addlinespace
    \multicolumn{7}{c}{\textit{True Classification Model Includes Covariates}} \\ 
    \addlinespace
    \multicolumn{7}{l}{WSWCs: 0.010; WSACs: 0.005; Cluster Sizes: $\mathcal{U}\{100,300\}$}\\
    \midrule
    Model 1 & 0.176 & -0.003 & 0.003 & 0.003 & 92.6\% & 93.9\% \\
    Model 2 & 0.176 & -0.062 & 0.003 & 0.003 & 76.0\% & 79.5\% \\
    \addlinespace
    \multicolumn{7}{l}{WSWCs: 0.010; WSACs: 0.005; Cluster Sizes: $\mathcal{U}\{500,1000\}$} \\
    \midrule
    Model 1 & 0.176 & -0.000 & 0.001 & 0.001 & 92.5\% & 94.0\% \\
    Model 2 & 0.176 & -0.060 & 0.001 & 0.001 & 43.6\% & 47.6\% \\
    \addlinespace
    \multicolumn{7}{l}{WSWCs: 0.100; WSACs: 0.050; Cluster Sizes: $\mathcal{U}\{100,300\}$} \\
    \midrule
    Model 1 & 0.165 & -0.003 & 0.005 & 0.004 & 92.1\% & 93.5\% \\
    Model 2 & 0.165 & -0.058 & 0.004 & 0.004 & 83.0\% & 85.4\% \\
    \addlinespace
    \multicolumn{7}{l}{WSWCs: 0.100; WSACs: 0.050; Cluster Sizes: $\mathcal{U}\{500,1000\}$} \\
    \midrule
    Model 1 & 0.165 & 0.000 & 0.003 & 0.002 & 92.4\% & 93.9\% \\
    Model 2 & 0.165 & -0.056 & 0.002 & 0.002 & 76.6\% & 79.5\% \\
    \bottomrule
\end{tabular}
    \label{tab:sim_tableapp2}
    \caption*{Model 1 - includes covariates in fitted classification model. Model 2 - does not include covariates in fitted classification model. Corrected coverage - coverage of $t$-intervals with $m-7$ degrees of freedom. WSAC refers to within-site across-clinician correlation. WSWC refers to within-site within-clinician correlation.}
\end{table}

\FloatBarrier

\subsection{Table S3}

The data generation follows (\ref{eq:dgm}) and (\ref{eq:paramst1}) from Table \ref{tab:sim_table1}. However, we add random intercepts $b_{i,y^*} \overset{\text{iid}}{\sim} N(0,\sigma^2_{b,y^*})$, where the variances were generated to match the ICC values of 0.01 and 0.1 as in the other two models. Specifically, we generate potential silver-standard outcomes as 
$$Y^*_{ij}(a) \sim \Bern\left\{\expit(\rho_{ij}(a,Y_{ij}(a),X_{ij},b_{i,y^*}) \right\}$$ with
$$\rho_{ij}(a,Y_{ij}(a),X_{ij},b_{i,y^*}) = \alpha_{a,y^*}+\beta_{a,y^*}^{T}X_{ij}+\delta_{a,y^*}Y_{ij}(a)+b_{i,y^*},$$
where all other parameters (other than for the new random effects) are as in (\ref{eq:paramst1}).

\begin{table}[ht]
    \centering
    \captionsetup{labelfont=bf,justification=raggedright,singlelinecheck=false}

    \caption{Evaluation of the small-sample performance of our proposed estimation method under differential misclassification with a non-random internal validation set. We generated data for 30 clusters as described in Section \ref{sec:simstudy} with reference Equations (\ref{eq:dgm}) and (\ref{eq:paramst1}) \textit{adding site random effects in the classification model generating silver-standard outcomes}. 8 scenarios are reported which have two levels of ICC=0.01 or 0.1, two levels of cluster sizes, $\mathcal{U}\{100,300\}$ or $\mathcal{U}\{500,100\}$, and exclusion or inclusion of covariates in the classification model generating silver-standard outcomes. Empirical bias and coverage of approximate 95\% normal intervals and small-sample corrected $t$-intervals using the cluster-robust (CR) sandwich variance are reported. The model-based CR variances are compared to the empirical variances for reference. Model 1 which accounts for covariates in fitting a classification model is juxtaposed against Model 2 which has no covariates, i.e., Model 2 estimates classification probabilities non-parametrically.}
\begin{tabular}{>{\centering\arraybackslash}p{1.5cm} 
                >{\centering\arraybackslash}p{1.5cm} 
                >{\centering\arraybackslash}p{1.5cm} 
                >{\centering\arraybackslash}p{1.5cm} 
                >{\centering\arraybackslash}p{1.5cm} 
                >{\centering\arraybackslash}p{1.5cm} 
                >{\centering\arraybackslash}p{1.5cm}}
    \toprule
    \textbf{\thead{Model}} & \textbf{\thead{True \\ ATE}} & \textbf{\thead{Empirical \\ Bias}} & \textbf{\thead{Empirical \\ Variance}} & \textbf{\thead{Model\\ Variance}} & \thead{\textbf{Coverage}} & \textbf{\thead{Corrected \\ Coverage}} \\
    \midrule
    \multicolumn{7}{c}{\textit{True Classification Model Excludes Covariates}} \\     
    \addlinespace
    \multicolumn{7}{l}{ICCs: 0.010; Cluster Sizes: $\mathcal{U}\{100,300\}$}\\
    \midrule
    Model 1 & 0.176 & -0.004 & 0.004 & 0.003 & 93.2\% & 94.5\% \\
    Model 2 &  0.176 & -0.001 & 0.003 & 0.003 & 93.0\% & 94.4\% \\
    \addlinespace
    \multicolumn{7}{l}{ICCs: 0.010; Cluster Sizes: $\mathcal{U}\{500,1000\}$} \\
    \midrule
    Model 1 & 0.176 & -0.001 & 0.001 & 0.001 & 93.2\% & 94.5\% \\
    Model 2 & 0.176 &  -0.000 & 0.001 & 0.001 & 92.8\% & 94.2\% \\
    \addlinespace
    \multicolumn{7}{l}{ICCs: 0.100; Cluster Sizes: $\mathcal{U}\{100,300\}$}\\
    \midrule
    Model 1 & 0.165 & -0.005 & 0.009 & 0.008 & 92.6\% & 94.0\% \\
    Model 2 & 0.165 & -0.002 & 0.007 & 0.007 & 92.1\% & 93.7\% \\
    \addlinespace
    \multicolumn{7}{l}{ICCs: 0.100; Cluster Sizes: $\mathcal{U}\{500,1000\}$} \\
    \midrule
    Model 1 & 0.165 & -0.002 & 0.006 & 0.005 & 93.0\% & 94.5\% \\
    Model 2 & 0.165 & -0.002 & 0.005 & 0.005 & 92.3\% & 93.8\% \\
    \bottomrule
    \addlinespace
    \multicolumn{7}{c}{\textit{True Classification Model Includes Covariates}} \\ 
    \addlinespace
    \multicolumn{7}{l}{ICCs: 0.010; Cluster Sizes: $\mathcal{U}\{100,300\}$}\\
    \midrule
    Model 1 & 0.176 & -0.002 & 0.004 & 0.003 & 93.0\% & 94.3\% \\
    Model 2 & 0.176 & -0.062 & 0.003 & 0.003 & 77.0\% & 80.3\% \\
    \addlinespace
    \multicolumn{7}{l}{ICCs: 0.010; Cluster Sizes: $\mathcal{U}\{500,1000\}$} \\
    \midrule
    Model 1 & 0.176 &  -0.000 & 0.001 & 0.001 & 92.7\% & 94.2\% \\
    Model 2 & 0.176 & -0.060 & 0.001 & 0.001 & 47.9\% & 51.8\% \\
    \addlinespace
    \multicolumn{7}{l}{ICCs: 0.100; Cluster Sizes: $\mathcal{U}\{100,300\}$} \\
    \midrule
    Model 1 & 0.165 & -0.003 & 0.009 & 0.008 & 93.0\% & 94.1\% \\
    Model 2 & 0.165 & -0.060 & 0.008 & 0.007 & 86.2\% & 88.2\% \\
    \addlinespace
    \multicolumn{7}{l}{ICCs: 0.100; Cluster Sizes: $\mathcal{U}\{500,1000\}$} \\
    \midrule
    Model 1 & 0.165 & -0.002 & 0.006 & 0.005 & 93.3\% & 94.5\% \\
    Model 2 & 0.165 & -0.060 & 0.005 & 0.005 & 83.4\% & 86.2\% \\
    \bottomrule
\end{tabular}
    \label{tab:sim_tableapp3}
    \caption*{Model 1 - includes covariates in fitted classification model. Model 2 - does not include covariates in fitted classification model. Corrected coverage - coverage of $t$-intervals with $m-7$ degrees of freedom. ICC - intracluster correlation coefficient.}
\end{table}

\FloatBarrier

\section{Supplemental Figure for ASPIRE}
\label{sec:suppfig}

\begin{figure}[H]
  \centering
  \captionsetup{labelfont=bf,justification=raggedright,singlelinecheck=false}
  \includegraphics[width=1\linewidth,scale = 1]{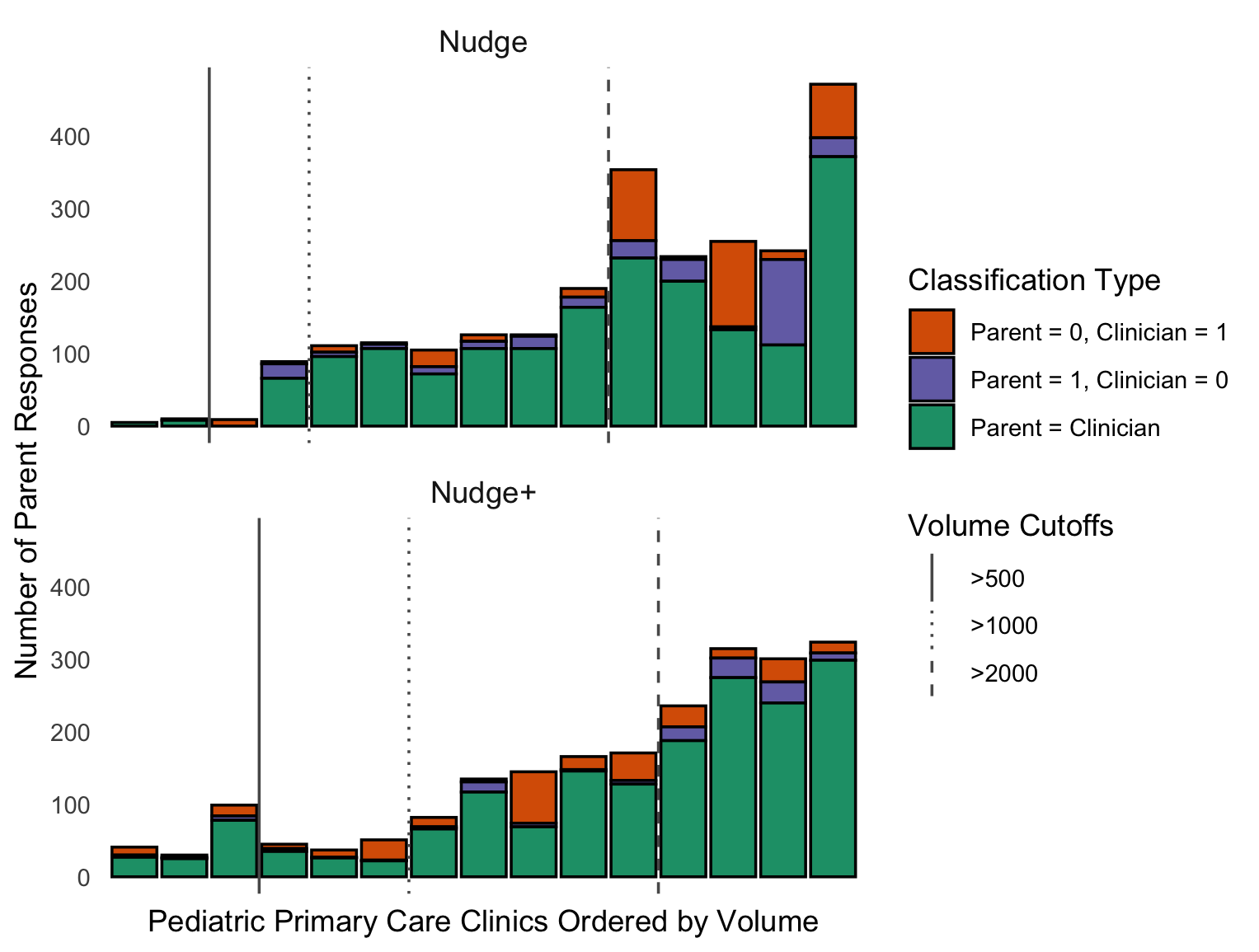}
  \centering
  \caption{This barplot displays classification types comparing gold-standard parent reports to silver-standard clinician reports in the validation data for each cluster in the ASPIRE study, stratified by treatment arm. The x-axis shows the pediatric primary care clinics ordered by clinic volume (measured as the number of visits in the year prior to the trial), and the y-axis represents the number of parent responses (i.e., number of collected gold-standard outcomes) following standard well-child visits. The subplots are indicated by the two treatment arms nudge and nudge+. The figure highlights heterogeneity in the amount of incorrect and correct classifications across clusters and treatments.}
  \label{fig:app1}
\end{figure}